\documentclass[prl,nofootinbib,superscriptaddress]{revtex4}
\usepackage[a4paper,left=1.5cm,right=1.5cm,top=3cm,bottom=3cm]{geometry}
\pdfoutput=1

\def\bma#1{\mbox{\boldmath{$#1$}}}
\usepackage{amsfonts,amssymb,mathrsfs,amsmath,esint}
\usepackage{slashed, cancel}
\usepackage{framed}
\usepackage{mdframed}
\usepackage{simplewick} % For Wick contractions
\allowdisplaybreaks

\usepackage{latexsym}
\usepackage{graphicx}
\graphicspath{{./Figures/}}
\usepackage[dvipsnames]{xcolor}
\usepackage{booktabs}
\usepackage{datetime}
\newdateformat{mydate}{\THEDAY{ }\monthname[\THEMONTH]{ }\THEYEAR}
\usepackage[toc]{appendix} %add option page to add a cover to the appendices

\usepackage{tikz}
\newcommand{\colboxed}[1]{#1}

\usepackage{hyperref}
\hypersetup{
  colorlinks,
  linktocpage=true,
  citecolor=rossos,
  linkcolor=bluscuro,
  urlcolor=bluchiaro,
  }
\definecolor{rossos}{cmyk}{0,1,1,0.55}
\definecolor{bluscuro}{rgb}{0.15, 0.2, .85}
\definecolor{bluchiaro}{cmyk}{1,.3,0.,0.1}

\newcommand{\arXiv}[2]{\href{http://arxiv.org/pdf/#1}{{\tt [#2/#1]}}}
\newcommand{\arXivold}[1]{\href{http://arxiv.org/pdf/#1}{{\tt [#1]}}}

\usepackage{bbold}%For identity-matrix symbol
\usepackage[utf8]{inputenc} %Support for additional characters in bibliography

\usepackage{feynmp}
\usepackage[
  font=footnotesize,
  labelfont=bf,
  margin=1cm,
  justification=raggedright,
]{caption} % Modifies captions

\newcommand{\abs}[1]{\left|#1\right|}
\def\lsim{\mathrel{\rlap{\lower4pt\hbox{\hskip0.5pt$\sim$}}
 \raise1pt\hbox{$<$}}}         %less than or approx. symbol
\def\gsim{\mathrel{\rlap{\lower4pt\hbox{\hskip0.5pt$\sim$}}
 \raise1pt\hbox{$>$}}}         %greater than or approx. symbol

\newcommand{\dd}{\mathrm{d}}
\newcommand{\etat}{\widetilde{\eta}}
\newcommand{\vx}{\mathbf{x}}
\newcommand{\vk}{\mathbf{k}}
\newcommand{\vp}{\mathbf{p}}

\newcommand{\calH}{\mathcal{H}}
\newcommand{\F}[1]{\widehat{#1}}
\newcommand{\FPsi}{\F{\Psi}}
\newcommand{\FS}{\F{\mathcal S}}
\newcommand{\Ic}{\mathcal{I}_c}
\newcommand{\Is}{\mathcal{I}_s}
\newcommand{\Ics}{\mathcal{I}_{c,s}}
\newcommand{\T}{\mathcal T}
\newcommand{\Pz}{\mathcal P_\zeta}
\newcommand{\Ph}{\mathcal P_h}
\newcommand{\Bh}{S_h}
\newcommand{\vdelta}{\delta^{(3)}}
\renewcommand{\S}{\mathscr S}
\newcommand{\OGW}{\Omega_\text{GW}}
\newcommand{\rhoGW}{\rho_\text{GW}}
\newcommand{\etaf}{\eta_f}
\newcommand{\trA}{0.8}
\newcommand{\trB}{0.65}

\newcommand{\etain}{\eta_\text{in}}

\usepackage[normalem]{ulem}

 \def\be   {\begin{equation}}   \def\ee   {\end{equation}}
 \def\ba   {\begin{array}}      \def\ea   {\end{array}}
 \def\bea  {\begin{eqnarray}}   \def\eea  {\end{eqnarray}}
 \def\bean {\begin{eqnarray*}}  \def\eean {\end{eqnarray*}}

\begin{document}

\title{A Cosmological Signature of the SM Higgs Instability: Gravitational Waves}
\author{J.~R.~Espinosa }
%\email{espinosa@ifae.es}
\address{Institut de F\'{\i}sica d'Altes Energies (IFAE), The Barcelona Institute of Science and Technology (BIST),
Campus UAB, 08193 Bellaterra, Barcelona, Spain}
\address{ICREA, Instituci\'o Catalana de Recerca i Estudis Avan\c{c}ats, \\ 
Passeig de Llu\'{\i}s Companys 23, 08010 Barcelona, Spain}
\address{Instituto de F\'isica Te\'orica IFT UAM/CSIC, \\ 
Calle Nicol\'as Cabrera 13. UAM, Cantoblanco. 28049 Madrid, Spain}
\author{D.~Racco }
%\email{davide.racco@unige.ch}
\author{A.~Riotto}
%\email{antonio.riotto@unige.ch}
\address{D\'epartement de Physique Th\'eorique and Centre for Astroparticle Physics (CAP), Universit\'e de Gen\`eve, 24 quai E. Ansermet, CH-1211 Geneva, Switzerland}

\date{\today}

\begin{abstract}
\noindent
A fundamental property of the  Standard Model  is that the Higgs potential  becomes unstable at large values of the Higgs field. For the current central values of the Higgs and top masses, the instability scale
is about $10^{11}$ GeV and therefore not accessible by colliders. We show that a possible signature of the Standard Model Higgs instability  is the production  of gravitational waves
sourced by  Higgs fluctuations generated during inflation. We fully characterise the   two-point correlator of such gravitational waves by computing its amplitude, the  frequency at peak,  the spectral index, as well as   their three-point correlators for various polarisations. We show that, depending on the Higgs and top masses, either  LISA or the  Einstein Telescope and Advanced-Ligo, could detect such stochastic background of gravitational waves. In this sense, collider and  gravitational wave physics  can provide fundamental and complementary informations. 
Since the mechanism described in this paper might also be responsible for the generation of dark matter under the form of primordial black holes,
this latter hypothesis may find its confirmation through the detection of gravitational waves.
\end{abstract}

\maketitle
\tableofcontents

%%%%%%%%%%%%%%%%%%%%%%%%%%%%%%%%%%%%%%%%%%
\section{I. Introduction and description of the scenario}
\noindent
The recent detection of gravitational  waves sourced  by a spiralling binary system made of two $\sim 30 M_\odot$ black holes \cite{ligo} has initiated the era of Gravitational Wave  (GW) cosmology \cite{revPBH} and  opened a new window to investigate the very  early stages of the evolution of the Universe \cite{revgw}. 
In particular, the Laser Interferometer Space Antenna (LISA) project  \cite{LISA},  as well as the Einstein Telescope (ET)  \cite{ET},  Advanced-Ligo \cite {AL}, and the Cosmic Explorer \cite{CE} at larger frequencies,  will search for  the stochastic gravitational wave background produced from different mechanisms, possibly identifying a primordial origin.

In this paper we point out  that a stochastic background of gravitational waves may be the probe of one of the most fundamental properties of the Standard Model (SM) of weak interactions:  the SM Higgs instability at high energies.
Within the SM,  the Higgs effective potential becomes deeper than the electroweak vacuum  for large values of the Higgs background field  when the   quartic Higgs coupling $\lambda$  becomes negative \cite{instab2,instab}. For instance, this happens (by choosing  the current central values of the Higgs and top masses) for Higgs field values of the order of $\Lambda\simeq 10^{11}$ GeV.
Despite this metastability condition of our present electroweak vacuum, its lifetime against decay  both via quantum  tunneling in flat spacetime or thermal fluctuations in the early Universe is by far longer than the age of the Universe \cite{instab,SSTU}. 

A natural and interesting question to ask is therefore which kind of physical phenomena might reveal, albeit indirectly, the presence of the SM instability. 
One option has been described in Ref. \cite{Espinosa:2017sgp} and   makes use of primordial inflation \cite{lrreview}, the early stage during  which the Universe expands exponentially and light
fields may be quantum mechanically excited. 
The dynamics  we consider can be conveniently divided in the following  stages:
\begin{enumerate}
\item In the first phase the Higgs has an initial value much smaller than the instability scale $\Lambda$. 
However, if it is lighter than the Hubble rate $H$, the classical  value  of the Higgs $h_{\rm c}$  field keeps receiving each Hubble time kicks of the order of $\pm (H/2\pi)$ and walks randomly.  
This dynamics is described by the stochastic equation \cite{espinosa}
\be
\ddot{h}_{\rm c}+3H\dot{h}_{\rm c}+V'(h_{\rm c})=3H\eta, ,\,\,\,\,V(h_{\rm c})\simeq -\frac{1}{4}\lambda h_{\rm c}^4,
\ee
where $\eta$ is  a Gaussian random noise with
\be
\langle\eta(t)\eta(t')\rangle=\frac{H^3}{4\pi^2}\delta(t-t').
\ee
If the Hubble rate is large enough, the Higgs field  can climb over the maximum of the potential  and roll down to the unstable region \cite{espinosa,cosmo2,Arttu,tetradis}.
\item In the second phase the Higgs has been pushed beyond the barrier of its effective potential and finds itself in the unbounded from below region. 
The motion starts being classically dominated over the quantum jumps at some time $t_*$ if the classical displacement, $(\Delta h)_{\rm cl}\simeq V'/(3H^2)$ in a Hubble time, is larger than the quantum jumps $(\Delta h)_{\rm q}\simeq H/(2\pi)$. 
This happens if the classical value of the Higgs $h_{\rm c}$ satisfies the relation
\be
h_{\rm c}^3 \gsim \frac{3H^3}{2\pi\lambda}.
\label{aa}
\ee
During this stage the classical equation of motion reads
\be
\ddot h_{\rm c}+3 H \dot h_{\rm c}+ V'(h_{\rm c})=0.
\label{a0}
\ee
For convenience, we will focus on those patches where classicality takes over during the last stages of inflation, say the last 20 $e$-folds or so. 
%Notice that these patches where the classical value of the Higgs satisfies the relation (\ref{aa}) are exponentially large as the correlation length where the Higgs field is homogeneous is of the order of $H^{-1} e^{H(t-t_i)}$, where $t_i$ indicates the beginning of inflation \cite{vilenkin}. 
At the beginning of this phase, the motion of the Higgs is friction dominated, 
$\ddot h_{\rm c}\lsim 3 H \dot h_{\rm c}$. This happens as long as $
 h_{\rm c}^2\lsim 3 H^2/\lambda$ and 
\be
\label{pp}
h_{\rm c}(t)\simeq \frac{h_*}{\left[1-2 \lambda h_*^2 (t-t_*)/3H\right]^{1/2}},
\ee
where  $t_*$ is the instant at which classicality takes over. 
When friction is subdominant, $h_{\rm c}$ rapidly increases
\be
\label{yx}
h_{\rm c}(t)\simeq \frac{\sqrt{2}}{\sqrt{\lambda}}\frac{1}{(t_{\rm p}-t)},
\ee
where $t_{\rm p}$ is the time when the Higgs hits the pole (see Appendix~A for more details).
%%
%\begin{framed}
%{\footnotesize 
%\noindent 
%To understand Eq. (\ref{yx}) one solves the equation
%\be
%\ddot{h}_{\rm c}-\lambda h_{\rm c}^3=0,
%\ee
%Taking the initial conditions $ h_{\rm c}(0)=h_0$ and $\dot{ h}_{\rm c}(0)=\dot h_0$, and using the fact that there is an integral of motion
%\be
%\frac{1}{2}\dot{h}_{\rm c}^2-\frac{\lambda}{4}h_{\rm c}^4=-E=\frac{1}{2}\dot{h}_0^2-\frac{\lambda}{4}h_0^4,
%\ee
%one finds the solution 
%\be
% h_{\rm c}(t)=h_0\alpha_0\,{\rm cn}\left(i\sqrt{\lambda}h_0\alpha_0\,t+{\rm cn}^{-1}(1/\alpha_0,1/2), \frac{1}{2}\right),
%\ee
%where ${\rm cn}(z,k)$ is one of the Jacobian elliptic functions
%and
%\be
%\alpha_0\equiv \left(1-\frac{2\dot{h}_0^2}{\lambda h_0^4}\right)^{1/4}.
%\ee
%%Using the identity ${\rm cn}(i z,k)= {\rm nc}(z,k')$, where $k'=\sqrt{1-k^2}$, we find 
%%\be
%% h_{\rm c}(t)=h_0\,{\rm nc}\left(\sqrt{\lambda}h_0\,t, \frac{\sqrt{3}}{2}\right),
%%\ee
%The function ${\rm cn}( ix,1/2)$ has poles at $x= K(1/2)$ with residue $-i\sqrt{2}$, where 
%\begin{equation}
%K(k)=\int_0^{\pi/2}\,\frac{{\rm d}\theta}{\sqrt{1-k\sin^2\theta}}.
%%\nonumber\\
%%K'(k)&=&K(k'),\nonumber\\
%%k'&=&\sqrt{1-k^2}.
%\end{equation}
%Around the pole the classical value of the Higgs can therefore be approximated by Eq.~\eqref{yx} with
%\be
%t_{\rm p}=\frac{1}{\sqrt{\lambda}h_0\alpha_0}
%\left[K(1/2)+i\,{\rm cn}^{-1}(1/\alpha_0,1/2)\right].
%\ee
%}
%\end{framed}
%%
\item The fast motion of the Higgs along the negative part of the potential may cause the patch of interest to experience an anti-de Sitter geometry which is potentially lethal. 
However, this situation can be rendered harmless and the patch end up to be our observed  Universe if the reheating temperature $T_{\rm RH}$ after inflation is large enough to push the Higgs back to our current vacuum \cite{tetradis,infdec,Espinosa:2017sgp}.
This happens because, thanks to the thermal interactions with the surrounding plasma, the Higgs potential is corrected to the form \cite{tetradis}
\be
V_T \simeq \frac{1}{2}m^2_T  h_{\rm c}^2,\,\,m_T^2\simeq 0.12\, T^2\,e^{-h_{\rm c}/(2\pi T)}.
\ee
If the reheating temperature is large enough, $T^2_{\rm RH }\gsim \lambda h_{\rm e}^2$, where $h_{\rm e}$ is the value of the Higgs when inflation ends, then the patch is rescued and the Higgs starts oscillating  (with a relativistic equation of state) around the current electroweak vacuum where it will settle after a while.\end{enumerate}
\begin{figure}[t]
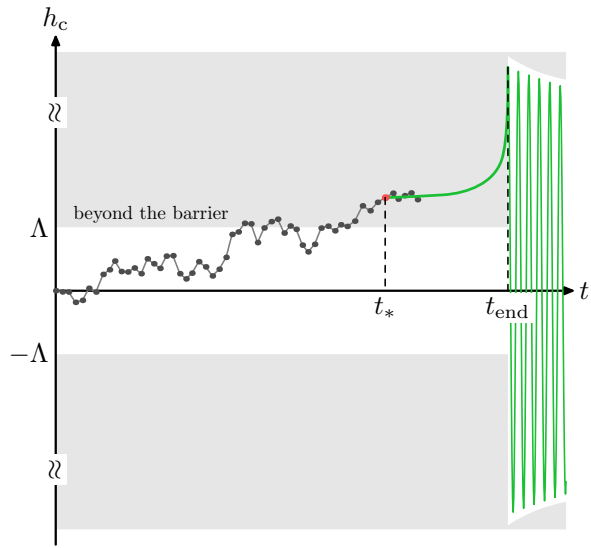

\raisebox{-\height}{$\hbox{ \convertMPtoPDF{./Figures/Higgs_Evolution.1}{0.85}{0.85} }$}
\caption{Evolution of the Higgs field background $h_\textrm{c}$  during inflation and reheating.}
\label{fig: evolution}
\end{figure}
Fig.~\ref{fig: evolution} summarises the dynamics of the classical value of the Higgs during the various stages.
Meanwhile, the perturbations of the Higgs field are excited and during inflation they satisfy the following equation of motion (in the flat gauge)
\be
\label{ay}
\delta\ddot{h}_{k}+3 H\delta\dot{h}_{k}+\frac{k^2}{a^2}\delta  h_{k}+V''(h_{\rm c})\delta  h_{k}=\frac{\delta  h_{k}}{a^3m_{\rm P}^2}\frac{{\rm d}}{{\rm d}t}\left(\frac{a^3}{H}\dot{h}_{\rm c}^2\right) ,
\ee
where $a$ is the scale factor, $m_{\rm P}$ is the reduced Planck mass, and the last term accounts for the backreaction of the metric perturbations. These perturbations are
born inside the Hubble radius with the standard Bunch-Davies vacuum, $\delta  h_{k}(k\gg aH)=(1/a\sqrt{2k})e^{-ik/aH}$ and, 
as soon as their physical wavelength becomes larger than  the Hubble radius, they rapidly grow driven by the rolling down of the 
Higgs field  \cite{Espinosa:2017sgp}
\be
\delta  h_{k}(k\ll aH)=\frac{H}{\sqrt{2 k^3}}\frac{\dot{h}_{\rm c}(t)}{\dot{h}_{\rm c}(t_k)}.
\ee
For the wavelengths leaving the Hubble radius the last 20 $e$-folds or so of inflation, the  gauge-invariant comoving curvature perturbation $\zeta(\vec x)$    is dominated by the Higgs perturbations  and reads 
\be
\zeta\simeq \frac{\dot{\rho}_h}{\dot\rho}\zeta_h,\,\,\,\, \zeta_h\simeq H\frac{\delta\rho_h}{\dot{\rho}_h} ,
\ee
(still in the flat gauge and we do not write down the subdominant standard component that is responsible for the cosmic microwave background anisotropies on much larger scales).
Since the Higgs energy-momentum tensor is separetely (covariantly) conserved during  inflation,   $\zeta_h$ is conserved on super-Hubble scales  and freezes in at the value  \cite{Espinosa:2017sgp}
\be
\dot{\zeta}_h(k\ll aH)=0\,\,\,{\rm and}\,\,\, \zeta_h(k\ll aH)=\frac{H^2}{\sqrt{2 k^3}\dot{h}_{\rm c}(t_k)},\,\,\,\, k=a(t_k)H.
\ee
After inflation ends, there is a fast  transient epoch  of  reheating during which the long wavelength Higgs perturbations are subject to  energy transfer involving the thermal plasma as their effective
mass suddenly jumps to its thermal value induced by the interactions with the plasma. 
After reheating is over the Higgs perturbations promptly decay  into radiation curvature perturbations which subsequently remain constant on super-Hubble scales as well (see Appendix C for a more detailed discussion).
Upon  Hubble reentry, if sizeable  enough, these perturbations source high peaks  in the matter power spectrum which collapse to form  Primordial Black Holes (PBHs).  
A first cosmological signature of the electroweak instability could be therefore  that the  dark matter (or a fraction thereof) is under the form of PBHs  seeded by Higgs fluctuations during inflation \cite{Espinosa:2017sgp}. 
In this scenario, no physics beyond the Standard Model should be invoked to explain the dark matter in our observed Universe, but anthropic arguments are necessary to explain the fine-tuning on the initial conditions.

In this paper we propose that a second signature of the SM instability might be a stochastic background  of gravitational waves  potentially detectable by the space-based interferometer LISA. 
Indeed, if there are large Higgs  perturbations generated during the last stages of inflation, responsible or not for the PBHs as dark matter, they inevitably act as a (second-order) source of primordial gravitational waves at horizon reentry. 
The goal of this paper is therefore to
\begin{enumerate}
\item  characterize the two-point correlator (power spectrum ${\cal P}_h$, its tilt as well as the frequency at the peak) of  gravitational waves induced by the first-order Higgs perturbations. Parametrically one expects ${\cal P}_h\sim {\cal P}_\zeta^2$ at Hubble crossing and therefore one can reach values of ${\cal P}_h$ as large as $10^{-4}$; the spectral tilt is also particularly interesting as the GW spectrum usually covers a large range of frequencies. 
 The study of the detectability of the spectral index of a generic GW background with  energy density  $\Omega_{\rm GW}(f)=A(f/f_*)^{n_T}$ can  be found in Ref. \cite{LISA}  as a function of  the frequency at the peak. For a signal peaked
at $f_*\sim 0.05$ Hz and $A\sim 10^{-12}$ one could constrain $n_T\lsim {\cal O}(1)$ and $n_T\gsim {\cal O}(7)$\footnote{For a frequency at the peak of 
$f_{\rm CMB}\sim 7.7\cdot 10^{-17}$ Hz, present CMB data already provide an upper bound on the amount of GWs, $\Omega_{\rm GW}^{\rm CMB}$, generated during inflation and one can write the GW energy density $\Omega_{\rm GW}=\Omega_{\rm GW}^{\rm CMB}(f/f_{\rm CMB})^{n_T}$, being $n_T$ the spectral tilt. A limit of $n_T\lsim 0.35$ can be obtained for the best LISA configuration with six links, five million km arm length and a five year mission \cite{LISA}.};

\item calculate the three-point correlator (bispectrum $B_h$) of the gravitational waves induced by the first-order Higgs perturbations. Parametrically one expects $B_h\sim {\cal P}_\zeta^3$ at Hubble crossing. 
Such a non-Gaussian signal, despite its presence in the primordial bispectrum, is unfortunately not observable today in a detector which collects signal simultaneously from all patches of the sky, as discussed in \cite{Bartolo:2018rku}.
\end{enumerate}
We will see that 

\begin{enumerate}

\item the energy density $\Omega_{\rm GW}$ of the GWs generated by the Higgs fluctuations is typically of the order of $10^{-8}$ at the peak. The latter is  reached  at frequencies ranging from  $10^{-2}$ to $10$ Hz.  
This should allow either  LISA or ET and Advanced-Ligo to detect the signal. 
Furthermore, as the frequency at the peak depends  sensitively on the Higgs and top mass, this will provide  complementary 
and fundamental information to be crossed with the ones provided  by  colliders with the possibility of either confirming or ruling out the origin of the GW signal; 

\item the  spectral index of the signal will have a characteristic behaviour: blue with $n_T\simeq 3$ for frequencies below the peak, and red with 
$n_T\simeq -0.6$ for frequencies above the peak frequency;

\item the bispectrum, in the case in which the two-point correlator is detectable by LISA, is mainly peaked in the so-called equilateral configurations. 
Summing up all polarisations we find the  characteristic consistency relation $(k_1k_2 k_3)^2 B_h\sim 3\cdot 10^4\, {\cal P}_h^{3/2}$.  
This non-Gaussian part of the GW signal is unfortunately unobservable in a GW detector, which unavoidably measures the sum of the signal from many uncorrelated patches of the sky. This sum is Gaussian by means of the Central Limit Theorem, and no bispectrum can be observed for any primordial GW background \cite{Bartolo:2018rku}.

\end{enumerate}
\noindent
The paper is organised as follows. In section II we describe the equation of motion for the GWs  and its solution; in section III we compute  the power spectrum of the GWs, while their  bispectrum  is calculated in section IV. Numerical results are found in section V and in section VI we conclude. The paper contains also various Appendices with extra useful material.

%%%%%%%%%%%%%%%%%%%%%%%%%%%%%%%%%%%%%%%%%%%%%%%%%%%%%%%%%%%%
\section{II. Equation of motion and its solution for Gravitational Waves}
\noindent
%%%%%%%%%%%%%%%%%%%%%%%%%%%%%%%%%%%%%%%%%%%%%%%%%%%%%%%%%%%%
Our goal is to evaluate the amount of gravitational waves produced during the radiation phase by the SM Higgs perturbations which in turn owe their origin to the previous period of inflation. 
The correct formalism to evaluate the contribution to the generation at second-order of tensor modes from first-order scalar perturbations has been  first discussed in \cite{Acquaviva:2002ud, Mollerach:2003nq, Ananda:2006af, Baumann:2007zm}. 
The first two parts of this section follow quite closely the notation of Appendix A of \cite{Ando:2017veq}.
Our convention for the signature of the metric is  $(-+++)$, so that the perturbed metric in the conformal Newtonian gauge reads
\begin{equation}
\mathrm ds^2=-a^2(1+2\Phi)\mathrm d\eta^2+a^2\left[(1-2\Psi)\delta_{ij}+\frac{1}{2}h_{ij}\right]\mathrm dx^i\mathrm dx^j ,
\end{equation}
where $\Phi$, $\Psi$ are the Bardeen potentials and the tensor perturbations $h_{ij}$ are transverse and traceless: $\partial_i h_{ij}=h_{ii}=0$.
In absence of anisotropy in the stress-energy tensor, we have $\Phi=\Psi$ (including  stress  gives only a small correction \cite{Baumann:2007zm}).  Furthermore, one can rewrite $h_{ij}$ in terms of the basis $\left\{e^{(+)}_{ij},e^{(\times)}_{ij}\right\}$ of polarisation tensors as follows
\begin{equation}
h_{ij}(\eta,\vx)=\int\frac{\mathrm d^3k}{(2\pi)^3}\left[h_{\vk}^{(+)}(\eta)e_{ij}^{(+)}(\vk)+h_{\vk}^{(\times)}(\eta)e_{ij}^{(\times)}(\vk)\right]e^{i\vk\cdot \vx}.
\end{equation}
The polarisation basis is given by
\begin{eqnarray}
e_{ij}^{(+)}(\vk)&=&\frac{1}{\sqrt 2}\left[e_i(\vk)e_j(\vk)-\bar e_i(\vk)\bar e_j(\vk)\right],\\
e_{ij}^{(\times)}(\vk)&=&\frac{1}{\sqrt 2}\left[e_i(\vk)\bar e_j(\vk)+\bar e_i(\vk) e_j(\vk)\right],
\end{eqnarray}
where $e_i(\vk)$ and $\bar e_i(\vk)$ are two three-dimensional vectors orthonormal to $\vk$, and the normalisation factor guarantees that $e_{ij}^{(+)}e_{ij}^{(+)}=e_{ij}^{(\times)}e_{ij}^{(\times)}=1$, $e_{ij}^{(+)}e_{ij}^{(\times)}=0$.

%%%%%%%%%%%%%%%%%%%%%%%%%%%%%%%%
\subsection{Equation of motion of GWs}
%%%%%%%%%%%%%%%%%%%%%%%%%%%%%%%
\noindent
The equation of motion for the GWs is obtained by extracting the tensor component of the Einstein equations expanded up to second order in perturbations
\begin{equation}
h_{ij}''+2\mathcal H h_{ij}'-\nabla^2 h_{ij}=-4 \mathcal T_{ij}{}^{lm}\mathcal S_{lm},
\label{eq: eom GW1}
\end{equation}
where $'$ denotes the derivative with respect to conformal time, $\mathcal H=a'/a$ is the conformal Hubble parameter, $\mathcal S_{lm}$ is the source term defined below in Eq.~\eqref{eq: source x space}. 
The projector $\mathcal T_{ij}{}^{lm}$ acting on the source term selects its transverse and traceless part. We define it in Fourier space (and use $\F{\ }\,$, when needed, to denote quantities in the conjugate space) as
\begin{equation}
\F{\mathcal T}_{ij}{}^{lm}(\vk)=e_{ij}^{(+)}(\vk)\,e^{(+)lm}(\vk)
  + e_{ij}^{(\times)}(\vk)\, e^{(\times)lm}(\vk).
\end{equation}
Our convention for the Fourier transform is the following:
\begin{equation}
\mathcal S_{lm}(\eta,\vx)=\int\frac{\mathrm d^3k}{(2\pi)^3}\FS_{lm}(\eta,\vk)e^{i\vk\cdot \vx} ,
\end{equation}
so that the equation of motion~\eqref{eq: eom GW1} reads, for each polarisation mode $s=(+),\,(\times)$,
\begin{equation}
{h_{\vk}^s}''(\eta) + 2 \mathcal H \, {h_{\vk}^s}'(\eta) + k^2 h_{\vk}^s(\eta)=\FS^s(\eta,\vk) ,
\label{eq: eom GW2}
\end{equation}
where $\FS^s(\eta,\vk)\equiv -4\,e^{s,lm}(\vk)\FS_{lm}(\eta,\vk)$.
The method of the Green function yields the solution
\begin{equation}
\colboxed{
h_{\vk}^s(\eta)=\frac{1}{a(\eta)} \int^\eta \dd \etat \,
  g_{\vk}(\eta, \etat) \, a(\etat) \, \FS^s(\etat,\vk),
} \label{eq: h solution}
\end{equation}
where the Green function $g_{\vk}(\eta,\etat)$ for a radiation-dominated (RD) Universe is
\begin{equation}
\colboxed{
g_{\vk}(\eta,\etat)=\frac{\sin\left[k(\eta-\etat)\right]}{k} \,\theta(\eta-\etat) ,
} \label{eq: Green function}
\end{equation}
$\theta$ being the Heaviside step function.

%%%%%%%%%%%%%%%%%%%%%%%%%%%%%%%%
\subsection{The source term for GWs}
%%%%%%%%%%%%%%%%%%%%%%%%%%%%%%%
\noindent
The source term $\FS_{ij}$ for GWs appearing in Eq.~\eqref{eq: eom GW1} arises at second order in the scalar perturbation $\Psi$ \cite{Acquaviva:2002ud}
\begin{equation}
\mathcal S_{ij}=4\Psi\partial_i\partial_j\Psi+2\partial_i\Psi\partial_j\Psi-\frac{4}{3(1+w)}\partial_i\left(\frac{\Psi'}{\mathcal H}+\Psi\right)\partial_j\left(\frac{\Psi'}{\mathcal H}+\Psi\right),
\label{eq: source x space}
\end{equation}
where $w$ is the equation of state of the fluid permeating the Universe at a given epoch.
Since the generation of GWs occurs mainly when the relevant modes re-enter the Hubble radius, which for the modes of our interest happens deeply into the RD era, we specialise to $w=1/3$. 
We rewrite the source in Fourier space, introducing
\begin{equation}
\FPsi(\eta,\vk)=\int\mathrm d^3x \, \Psi(\eta,\vx) \,e^{-i\vk\cdot\vx}
\end{equation}
so the right hand side of Eq.~\eqref{eq: eom GW2} becomes (we omit the temporal dependence for brevity)
\begin{equation}
\FS^s(\eta,\vk)= 4 \int\frac{\dd^3p}{(2\pi)^3}e^{s,ij}(\vk) p_i p_j
  \left[2\FPsi(\vp)\FPsi(\vk-\vp)
    + \left(\FPsi(\vp) +\frac 1\calH \FPsi'(\vp) \right) 
    \left(\FPsi(\vk -\vp) +\frac 1\calH \FPsi'(\vk -\vp) \right) \right].
\label{eq: source}
\end{equation}
The expression inside squared brackets is explicitly symmetric under the exchange of $\vp$ and $\vk -\vp$.

The scalar perturbation $\Psi(\eta,\vk)$ is directly related to the gauge invariant comoving curvature perturbation by $\Psi = \tfrac 23 \zeta$ \cite{lrreview}. 
We define then the transfer function $T(\eta,k)$ through the relation
\begin{equation}
\FPsi(\eta,\vk)=\frac 23 T(\eta,k) \zeta(\vk) ,
\label{eq: Psi to zeta}
\end{equation}
and its expression is given in the RD era by  
\begin{equation}
\colboxed{
T(\eta,k)=\T(k\eta), \quad  
\T(z)= \frac{9}{z^2}\left[ \frac{\sin (z/\sqrt 3)}{z/\sqrt 3} -\cos(z/\sqrt 3) \right]. 
}
\label{eq: transfer}
\end{equation}
We can rewrite the source term~\eqref{eq: source} as
\begin{equation}
\colboxed{
\FS^s(\eta,\vk)=\frac 49 \int\frac{\dd^3 p}{(2\pi)^3}\, e^{s}(\vk,\vp)f(p,|\vk-\vp|,\eta)\, \zeta(\vp)\zeta(\vk-\vp),
} \label{eq: source zeta}
\end{equation}
where we have introduced
\begin{equation}
\colboxed{
e^{s}(\vk,\vp)\equiv e^{s,ij}(\vk)p_i p_j=
\begin{cases}
\frac{1}{\sqrt 2}p^2\sin^2\theta\cos 2\phi & \textrm{for } s=(+),\\
\frac{1}{\sqrt 2}p^2\sin^2\theta\sin 2\phi & \textrm{for } s=(\times),
\end{cases}
}\label{eq: polarisation tensors spherical}
\end{equation}
where $(p,\theta,\phi)$ are the coordinates of $\vp$ in a spherical coordinate system whose $(\hat x, \hat y, \hat z)$ axes are aligned with $(e(\vk), \bar e(\vk),\vk$), and
\begin{equation}
\colboxed{
f(k_1,k_2,\eta) \equiv 4 \left[ 2 T(\eta, k_1)T(\eta, k_2)	+ 
  \left( T(\eta, k_1)+ \frac{1}{\calH} T'(\eta, k_1)\right) 
  \left( T(\eta, k_2)+ \frac{1}{\calH} T'(\eta, k_2)\right) \right].
  }
\label{eq: f}
\end{equation}

%%%%%%%%%%%%%%%%%%%%%%%%%%%%%%%%%%%%%%%%%%
\subsection{A compact expression for GWs with a numerical integration over time}
%%%%%%%%%%%%%%%%%%%%%%%%%%%%%%%%%%%%%%%%%%
Let us rewrite the solution for the GWs $h_\vk^s(\eta)$ by collecting the results of \eqref{eq: h solution}, \eqref{eq: Green function}, \eqref{eq: source zeta}
\begin{align}
h_\vk^s(\eta) = & \frac{1}{a(\eta)} \int^\eta \dd\etat \frac{\sin(k\eta) \cos(k\etat) - \cos(k\eta) \sin(k\etat)}{k} a(\etat)\,\frac 49 \int \frac{\dd^3p}{(2\pi)^3} e^s(\vk,\vp) f(p,|\vk-\vp|, \etat) \zeta(\vp)\zeta(\vk-\vp) = \nonumber\\
= & 
\frac 49 \int \frac{\dd^3p}{(2\pi)^3} \frac{1}{k^3 \eta} e^s(\vk,\vp)  \zeta(\vp)\zeta(\vk-\vp) \left[ 
\int^\eta k\dd\etat \, (k\etat)\, \Big(\sin(k\eta) \cos(k\etat) - \cos(k\eta) \sin(k\etat)\Big) f(p,|\vk-\vp|, \etat)\right],
\label{eq: h rewritten}
\end{align}
where we have expressed the scale factor in terms of conformal time during RD, $a(\etat)/a(\eta)=\etat/\eta$.

The tensor modes begin to be generated at the time at which the wavelength  $1/k$ re-enters the comoving Hubble radius, given that the source is damped on super-Hubble scales. We provide in Appendix~D the analytical results both for a generic lower extreme of integration $\etain$ and for two values which are equivalent in practice: $\etain=0$ (strictly speaking, the exact result) and $\etain=k^{-1}$, which is the value that is chosen for the numerical results in the remainder of the paper.
The transfer function \eqref{eq: transfer} decays as $\eta^{-2}$, so that the generation of tensor modes is completed within a time which is a few orders of magnitude larger than $k^{-1}$, around $\eta \sim \mathcal O (10^3)k^{-1}$.
Therefore the extrema of the integral over $\etat$ in Eq.~\eqref{eq: h rewritten} are $\etat=\etain$ and the current time $\eta \gg \mathcal O(10^3)k^{-1}$, so that we can approximate it to $\etat \to \infty$.

\begin{figure}[t]
\centering
\includegraphics[width=.5\textwidth]{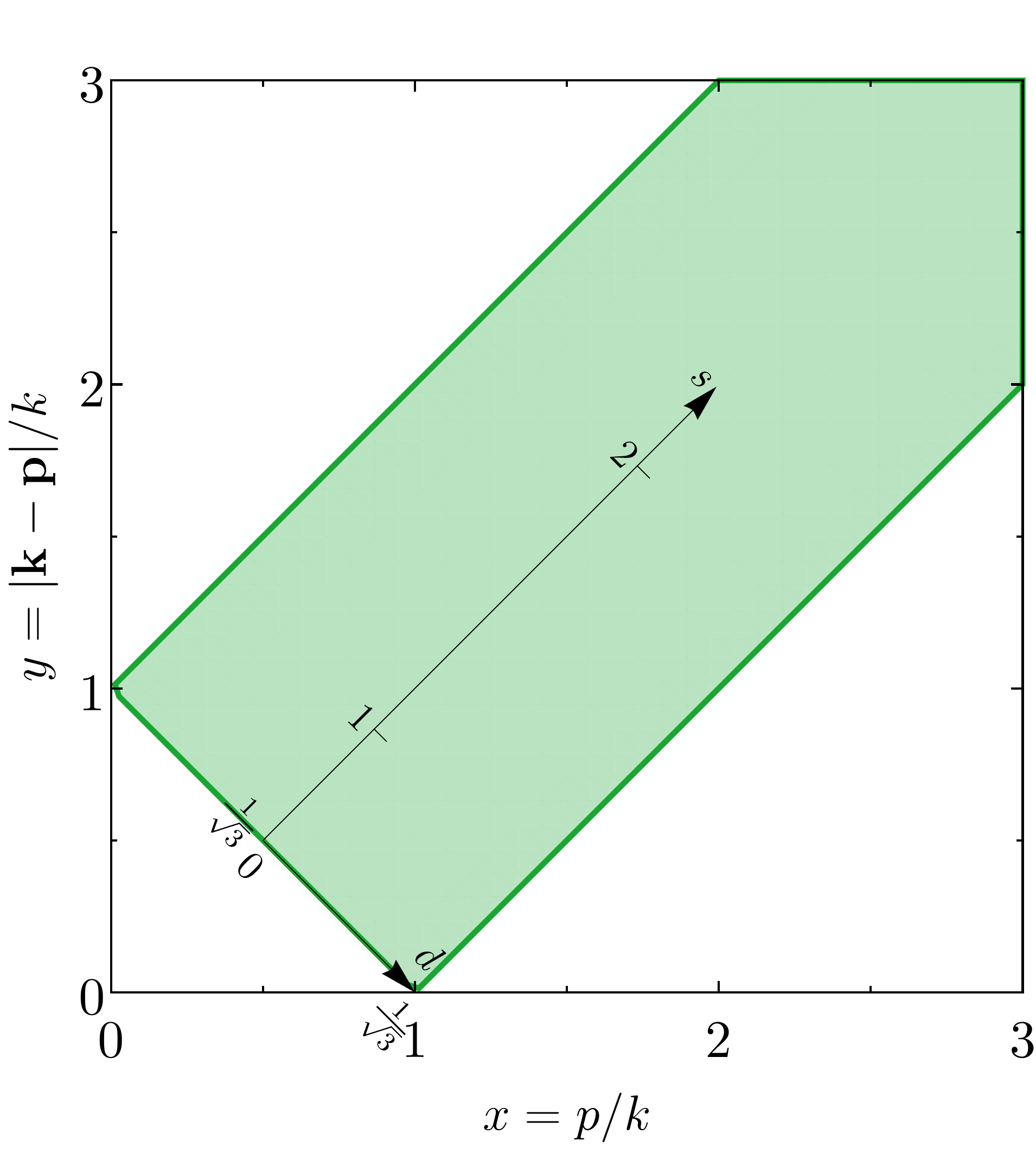}
\caption{Domain for the variables $x=p/k, \, y=|\vk-\vp|/k$ allowed by the triangular inequality, superimposed with the $(d,s)$ coordinates defined in Eq.~\eqref{eq: xy to ds}.}
\label{fig: xy to ds}
\end{figure}

\begin{figure}[h!]
\centering
\includegraphics[width=.49\textwidth]{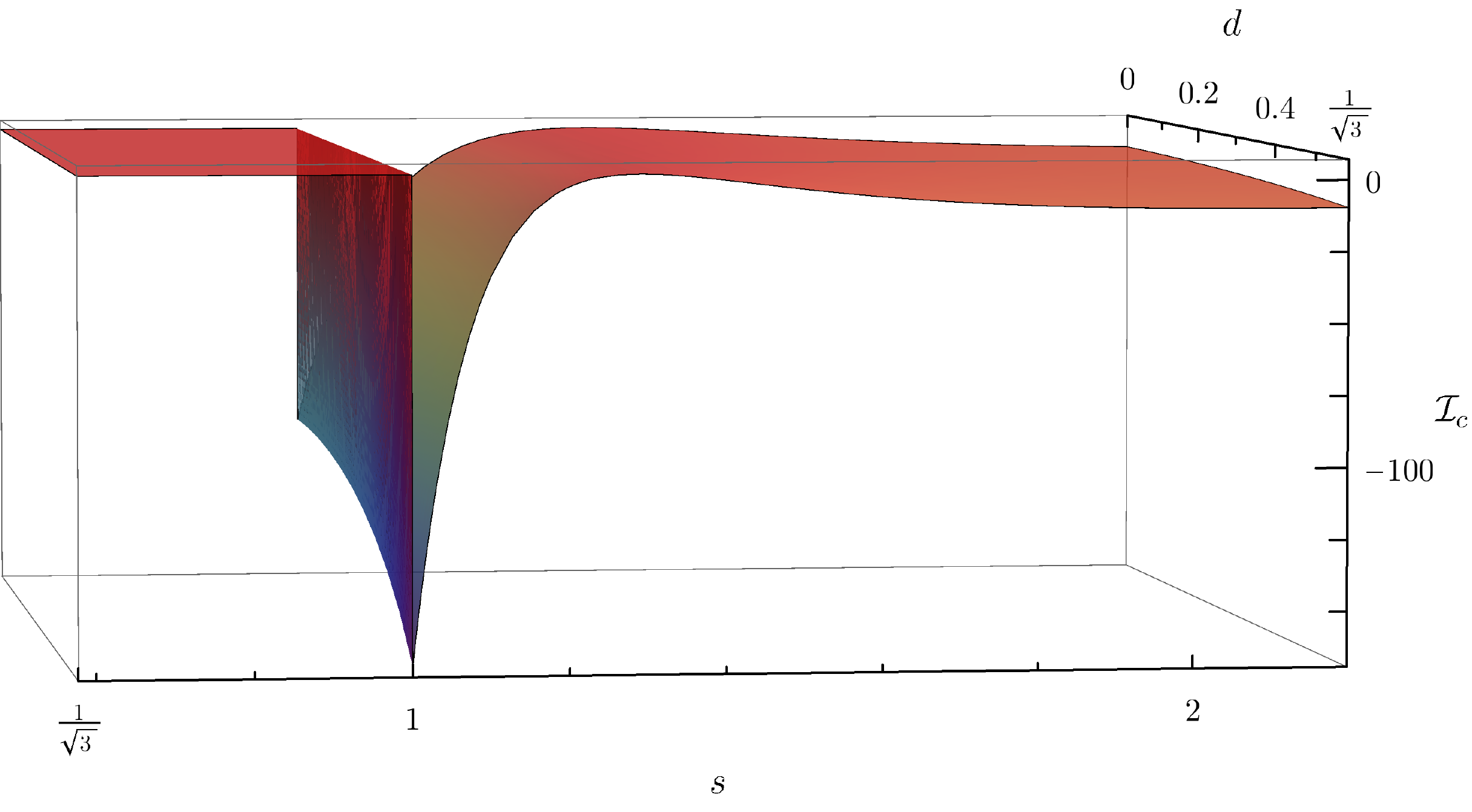}
\includegraphics[width=.49\textwidth]{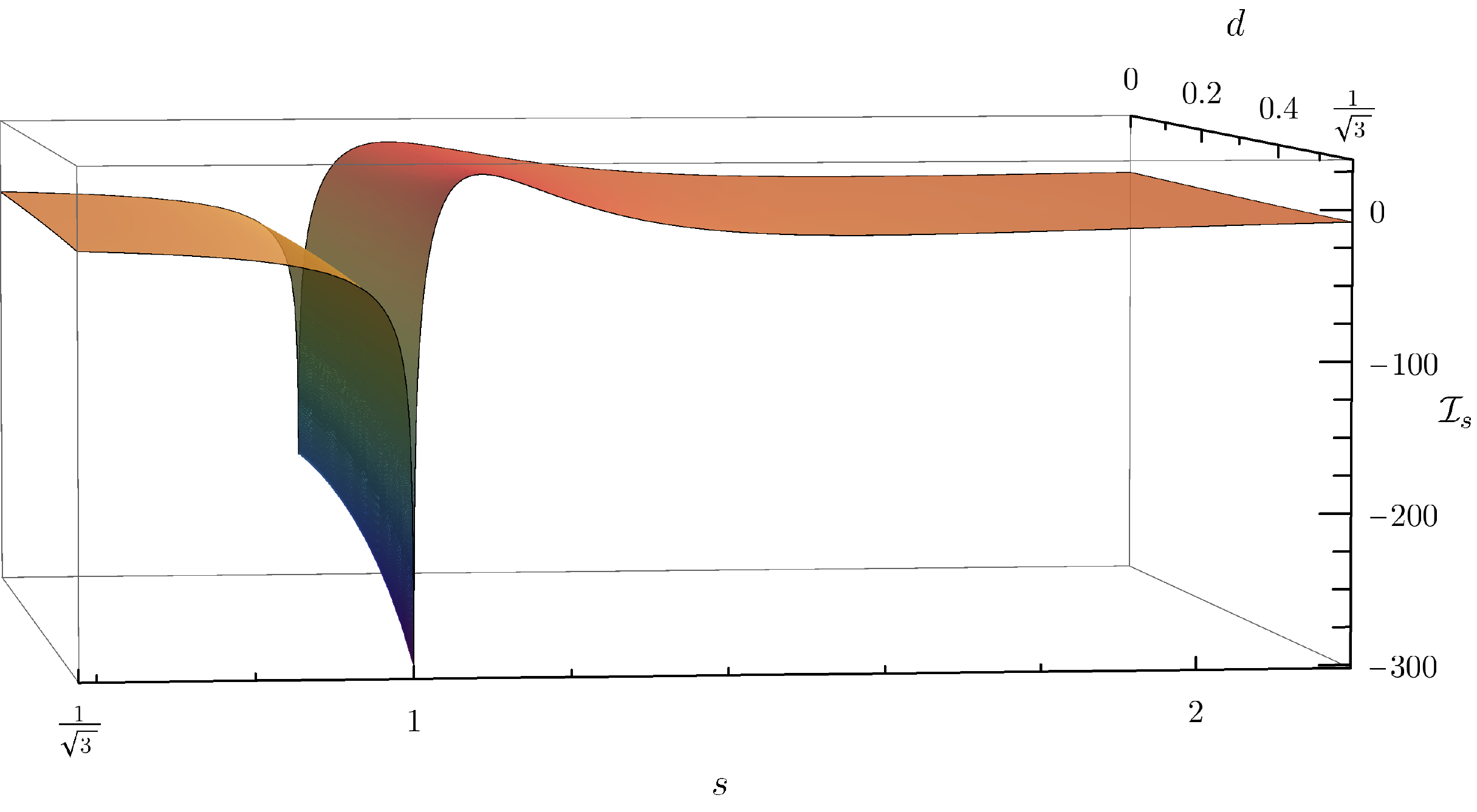}
\caption{3D plots of $\Ic$ (\textit{left} plot) and $\Is$ (\textit{right} plot), defined in Eq.~\eqref{eq: Ic, Is}, as a function of $(d,s)$ (Eq.~\eqref{eq: xy to ds}).}
\label{fig: integral eta 3D}
\end{figure}

\begin{figure}[h!]
\centering
\includegraphics[width=.6\textwidth]{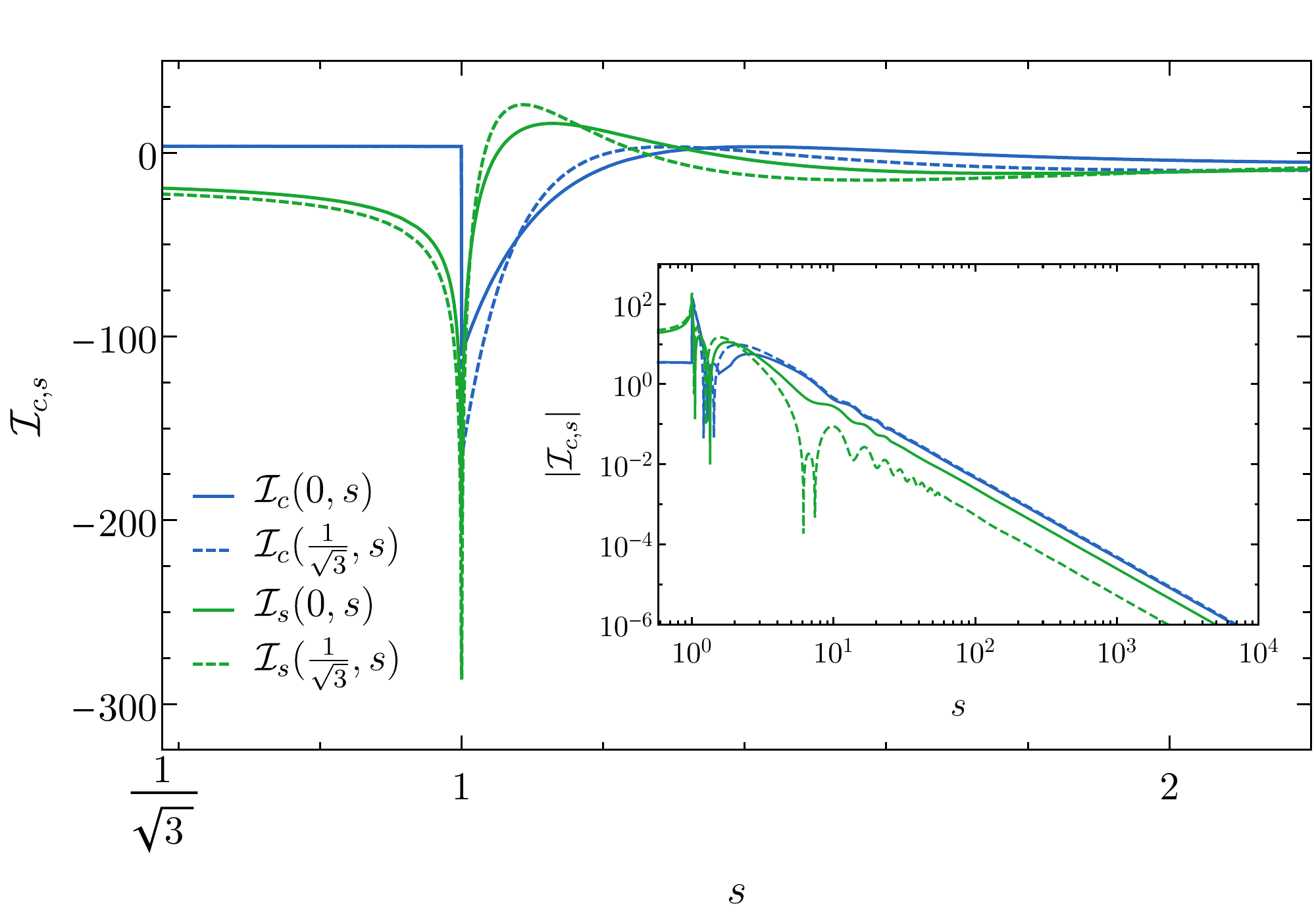}
\caption{Behaviour of the integrals $\Ic$, $\Is$, defined in Eq.~\eqref{eq: Ic, Is}, as a function of $s$ (Eq.~\eqref{eq: xy to ds}), for the two extremal values of $d=|x-y|/\sqrt 3$.}
\label{fig: integral eta 2D}
\end{figure}

The dimensionless expression contained in square brackets in Eq.~\eqref{eq: h rewritten} can be computed analytically, in order to facilitate the calculation of the two- and three-point functions. 
We denote
\begin{equation}
x=\frac pk , \qquad y = \frac{|\vk-\vp|}{k},
\end{equation}
and we use the dimensionless time variable $\tau \equiv k\etat$, and we input the Hubble rate $\calH =aH=\eta^{-1}$ during RD. We can then rewrite Eq.~\eqref{eq: h rewritten} as
\begin{equation}
\colboxed{
h_\vk^s(\eta) = \frac 49 \int \frac{\dd^3p}{(2\pi)^3} \frac{1}{k^3\eta} e^s(\vk,\vp)  \zeta(\vp)\zeta(\vk-\vp) \Big[ 
\Ic(x,y) \cos(k\eta) + \Is(x,y) \sin(k\eta)\Big],}
\label{eq: h with Ic, Is}
\end{equation}
where we have introduced two functions, $\Ic$ and  $\Is$, that can be computed analytically (the reader can find the analytical result in Appendix D) 
\begin{equation}
\begin{aligned}
\Ic(x,y) &= \int_{\etain}^\infty \dd\tau \, \tau (-\sin \tau) \cdot 4 \Big\{ 2\T(x\tau)\T(y\tau) + \Big[\T(x\tau) + x\tau\, \T'(x\tau) \Big] \Big[\T(y\tau) + y\tau\, \T'(y\tau) \Big] \Big\}, \\
\Is(x,y) &= \int_{\etain}^\infty \dd\tau \, \tau (\cos \tau) \cdot 4 \Big\{ 2\T(x\tau)\T(y\tau) + \Big[\T(x\tau) + x\tau\, \T'(x\tau) \Big] \Big[\T(y\tau) + y\tau\, \T'(y\tau) \Big] \Big\}.
\end{aligned}
\label{eq: Ic, Is} 
\end{equation}
The domain in the $(x,y)$ plane is shown in the left plot of Fig.~\ref{fig: xy to ds}: it consists of the configurations allowed by the triangular inequality applied to the triangle formed by the vectors $\vk,\, \vp,\, \vk-\vp$, and is given by 
\begin{equation}
(x+y\geq 1) \wedge (x+1\geq y) \wedge (y+1\geq x). 
\label{eq: domain xy}
\end{equation}
It is useful to introduce two auxiliary variables $(d,s)$ in terms of $(x,y)$, which simplify the expression of $\Ic$, $\Is$ for the purpose of an analytical integration,
\begin{equation}
d=\frac{1}{\sqrt{3}}|x-y|, \qquad  s=\frac{1}{\sqrt{3}}(x+y) ,\qquad  (d,s) \in [0,1/\sqrt{3}]\times[1/\sqrt{3},+\infty)
\label{eq: xy to ds}
\end{equation}
This redefinition of domain is illustrated in Fig.~\ref{fig: xy to ds}.
\newline
The result for the analytical calculation of the integrals $\Ic$, $\Is$ for each point $(d,s)$ is shown in Figs.~\ref{fig: integral eta 3D} and \ref{fig: integral eta 2D}.
\newline
We observe that the numerical value of $\Ic(d,s)$, $\Is(d,s)$ is nearly independent of $d=|x-y|/\sqrt{3}$. 
More interestingly, the integrals $\Ic$, $\Is$ are spiked for a value of $s\sim 1$ corresponding to $p+|\vk -\vp| \sim \sqrt 3 k$. 
The reason for this is that the integrands of $\Ic$ and $\Is$ are products of trigonometric functions of $\tau$ times a rational function of $\tau$, and the oscillating behaviour determines cancellations in the final result. 
Only for $p+|\vk -\vp| \sim \sqrt 3 k$ there appear some terms in the integrand with the square of a trigonometric function and thus with a definite sign, and this increases the final result. 
Notice that the factor $\sqrt 3$ is simply due to the factor $\sqrt w$ appearing in the arguments of the transfer function of Eq.~\eqref{eq: transfer}, and not to geometrical reasons.

%%%%%%%%%%%%%%%%%%%%%%%%%%%%%%%%%%%%%%%%%%%%%%%%%%%%%%%%%%%%
\section{III. The power Spectrum of Gravitational Waves}
%%%%%%%%%%%%%%%%%%%%%%%%%%%%%%%%%%%%%%%%%%%%%%%%%%%%%%%%%%%%
\noindent
In this section we present the generic derivation of the two-point function and the power spectrum of gravitational waves. 
This result has been already derived and exposed in Refs. \cite{Acquaviva:2002ud, Mollerach:2003nq, Ananda:2006af, Baumann:2007zm}. 
The goal of the present section is to match  it  with  our notation, and to prepare an analogous derivation of the three-point function of GWs in the next section.
In section~V we will use the formul\ae\ obtained here  to calculate the power spectrum and the three-point function of GWs generated in our scenario.

%%%%%%%%%%%%%%%%%%%%%%%%%%%%%%%%%
\subsection{Two-point function of GWs}
%%%%%%%%%%%%%%%%%%%%%%%%%%%%%%%%%
\noindent
We begin by writing the definition of two-point function, with the use of Eq.~\eqref{eq: h with Ic, Is}
\begin{equation}
\begin{aligned}
\langle h^r(\eta, \vk_1) h^s(\eta, \vk_2) \rangle = & 
  \left(\frac 49 \right)^2 \int \frac{\dd^3 p_1}{(2\pi)^3} \int \frac{\dd^3 p_2}{(2\pi)^3} 
  \frac{1}{k_1^3 k_2^3 \eta^2} e^r(\vk_1,\vp_1) e^s(\vk_2,\vp_2) 
  \Big\langle \zeta(\vp_1) \zeta(\vk_1-\vp_1) \zeta(\vp_2) \zeta(\vk_2-\vp_2) \Big\rangle \cdot \\
  & \cdot 
  \big[ \cos(k_1\eta) \Ic(x_1,y_1) + \sin(k_1\eta) \Is(x_1,y_1)\big] 
  \big[ \cos(k_2\eta) \Ic(x_2,y_2) + \sin(k_2\eta) \Is(x_2,y_2)\big] ,
\end{aligned}
\label{eq: two-pt h def}
\end{equation}
where $x_i=p_i/k_i$, $y_i=|\vk_i- \vp_i|/k_i$.
To evaluate the four-point function of the curvature perturbation $\zeta$ we proceed as usual, noting that at leading order it is a Gaussian variable defined by the dimensionless power spectrum $\Pz$
\begin{equation}
\left\langle \zeta(\vk_1)\zeta(\vk_2)\right\rangle \equiv (2\pi)^3 \vdelta(\vk_1+\vk_2)\frac{2\pi^2}{k_1^3} \Pz(k_1) ,
\label{eq: def P zeta}
\end{equation}
and the four-point function of $\zeta$ of the first line of \eqref{eq: two-pt h def} has two possible non-vanishing contractions for $\vk_1, \vk_2\neq 0$. The two contributions give the same result, given that they correspond to each other up to a shift $\vp_2\to (\vk_2-\vp_2)$, which is a symmetry of Eq.~\eqref{eq: two-pt h def}, see Appendix B for details.
We can evaluate then Eq.~\eqref{eq: two-pt h def} for any of the two configurations, and multiply the final result by 2. After integrating over $\vp_2$  one gets
\begin{multline}
\langle h^r(\eta, \vk_1) h^s(\eta, \vk_2) \rangle =  
  (2\pi)^3 \vdelta(\vk_1+\vk_2)\cdot 2 \left(\frac 49 \right)^2
  \int \frac{\dd^3 p_1}{(2\pi)^3}
  \frac{1}{k_1^6 \eta^2} e^r(\vk_1,\vp_1) e^s(\vk_1,\vp_1) 
  \frac{2\pi^2}{p_1^3} \frac{2\pi^2}{|\vk_1-\vp_1|^3} \cdot \\
  \cdot \Pz(p_1) \Pz(|\vk_1-\vp_1|) 
  \big[ \cos^2(k_1\eta) \Ic(x_1,y_1)^2 
  + \sin^2(k_1\eta) \Is(x_1,y_1)^2 
  + \sin(2k_1\eta) \Ic(x_1,y_1)\Is(x_1,y_1) \big].
\label{eq: two-pt h contracted}
\end{multline}
Let us refer to a system of spherical coordinates $(p_1,\theta,\phi)$ oriented around the axis $\vk_1$, and denote $x\equiv x_1=p_1/k_1$, $y\equiv y_1=|\vk_1-\vp_1|/k_1$. In these variables one has $\vp_1= (k_1 x, \cos^{-1}\left((1+x^2-y^2)/2x\right),\phi)$. We perform the following change of integration variables
\begin{equation}
\int\dd^3p_1 \longrightarrow k_1^3 \iint_\S \dd x\, \dd y\, x^2 \frac yx \int_0^{2\pi} \dd \phi,
\end{equation}
where $\S$ is the infinite strip shown in Fig.~\ref{fig: xy to ds}.
The integral over $\phi$ can be easily solved analytically, and selects only some of the possible couples of polarisations $(r,s)$ to give a non-vanishing result. With the use of Eq.~\eqref{eq: polarisation tensors spherical} we obtain
\begin{equation}
\int_0^{2\pi}\dd\phi \,e^r(\vk_1,\vp_1) e^s(\vk_1,\vp_1) = \frac{k_1^4}{2} x^4 \left[1-\frac{(1+x^2-y^2)^2}{4x^2} \right]^2\cdot \pi \,\delta^{rs}.
\label{eq: polarisation 2-pt parity}
\end{equation}
By collecting the results of the last three equations we get the final expression for the two-point function of GWs:
\begin{multline}
\langle h^r(\eta, \vk_1) h^s(\eta, \vk_2) \rangle =  
  (2\pi)^3 \vdelta(\vk_1+\vk_2) \frac{2\pi^2}{k_1^3} \delta^{rs}\cdot 
  2 \left(\frac 49 \right)^2 \frac{1}{k_1^2\eta^2}
  \iint_\S \dd x\,\dd y
  \frac{x^2}{8y^2} \left[1-\frac{(1+x^2-y^2)^2}{4x^2} \right]^2
  \cdot \\
  \cdot \Pz(k_1x) \Pz(k_1y) 
  \big[ \cos^2(k_1\eta) \Ic(x,y)^2 
  + \sin^2(k_1\eta) \Is(x,y)^2 
  + \sin(2k_1\eta) \Ic(x,y)\Is(x,y) \big].
\label{eq: two-pt h final}
\end{multline}
The integrand is explicitly symmetric under exchange of $x$ and $y$. 
From Eq.~\eqref{eq: two-pt h final} and the definition of the power spectrum of GWs
\begin{equation}
\left\langle h^r(\eta,\vk_1) h^s(\eta,\vk_2)\right\rangle \equiv (2\pi)^3 \vdelta(\vk_1+\vk_2)\, \delta^{rs}\,\frac{2\pi^2}{k_1^3} \Ph(k_1)
\label{eq: def P h}
\end{equation}
we can extract $\Ph(\eta,k)$:
\begin{equation}
\hspace{-1em}
\colboxed{
\Ph(\eta,k) =  
  \frac{4}{81}\frac{1}{k^2\eta^2}
  \iint_\S \dd x\,\dd y
  \frac{x^2}{y^2} \left[1-\frac{(1+x^2-y^2)^2}{4x^2} \right]^2
  \Pz(kx) \Pz(ky) 
  \big[ \cos^2(k\eta) \Ic^2 
  + \sin^2(k\eta) \Is^2 
  + \sin(2k\eta) \Ic\Is \big],
}
\label{eq: PS GWs}
\end{equation}
where for brevity we do not write the arguments of the functions $\Ic(x,y)$ and $\Is(x,y)$, defined in Eq.~\eqref{eq: Ic, Is} and plotted in Figs.~\ref{fig: integral eta 3D} and \ref{fig: integral eta 2D}.

%%%%%%%%%%%%%%%%%%%%%%%%%%%%%%%%%%%%%%%%%%
\subsection{The energy density of GWs}
%%%%%%%%%%%%%%%%%%%%%%%%%%%%%%%%%%%%%%%%%%
\noindent
In this section we derive the expression for the energy density  of GWs, and its fraction $\OGW$ relative to the critical energy density.
The energy density of GWs is \cite{Maggiore:1999vm}
\begin{equation}
\rhoGW(\eta,\vx) = \frac{m_P^2}{16 a^2(\eta)}
  \left\langle \frac 12 \overline{\left(h_{ij}'\right)^2} + \frac 12 \overline{ \left(\nabla h_{ij} \right)^2} \right\rangle
  \simeq\frac{m_P^2}{16 a^2(\eta)} \left<\overline{\left(\nabla h_{ij}\right)^2}\right> ,
\end{equation}
where the overlines denote an average over time. 
This expression for the energy density can be rewritten in terms of the power spectrum of GWs as follows
\begin{align}
\rhoGW(\eta) & =\int \dd \ln k\, \rhoGW(\eta,k), \\
\rhoGW(\eta,k) & = \frac{m_P^2}{8} \left( \frac{k}{a(\eta)}\right)^2 \overline{\Ph(\eta,k)}.
\label{eq: rho GW}
\end{align}
We can then define the density parameter of GWs per logarithmic interval of $k$,
\begin{equation}
\OGW(\eta,k) = \frac{\rhoGW(\eta,k)}{\rho_\text{cr}(\eta)}=\frac{1}{24} \left(\frac{k}{\calH(\eta)}\right)^2 \overline{\Ph(\eta,k)}.
\end{equation}
The expression for the power spectrum that we have computed in the previous section holds only during the RD era. 
The energy density of GWs decays as radiation, so we can easily estimate the fraction of energy density of GWs in terms of the current energy density of radiation $\Omega_{r,0}$ and $\OGW(\etaf,k)$ at a generic time $\etaf$ during the RD era, before the SM degrees of freedom become non relativistic.
Taking $\etaf$ at a time when the top quark is non relativistic, and assuming that there are no extra relativistic degrees on freedom on top of the SM value $g_{*,f}=106.75$, the radiation density at $\etaf$ is related to its current value by the conservation of entropy:
\begin{equation}
c_g \equiv \frac{a_f^4 \rho_{r,f}}{a_0^4\rho_{r,0}} = \frac{g_{*,f}}{g_{*,0}} \left(  \frac{g_{*S,0}}{g_{*S,f}}\right)^{4/3} \approx 0.4 \,.
\end{equation}
We can then write the current energy density of GWs by rescaling it from $\etaf$ until today by $a^{-4}$:
\begin{equation}
\OGW(\eta_0,k) = \left(c_g\Omega_{r,0}\right)\, \OGW(\etaf,k) = 
c_g \frac{\Omega_{r,0}}{24} \frac{k^2}{\calH (\etaf)^2} \overline{\Ph(\etaf,k)}.
\label{eq: Omega GW}
\end{equation}
We can collect the results of Eqs.~\eqref{eq: PS GWs} and \eqref{eq: Omega GW}, plug $\calH(\etaf) = 1/\etaf$ (valid through RD until $\etaf$), and perform a simplification for the average over time justified by the fact that $k\eta\gg 1$
\begin{equation}
\overline{\frac{\cos^2(k\eta)}{\eta^2}} \sim 
\overline{\frac{\sin^2(k\eta)}{\eta^2}} \sim \frac 12 \frac{1}{\eta^2} , \quad
\overline{\frac{\sin(2k\eta)}{\eta^2}} \sim 0. 
\end{equation}
We finally obtain  the current energy density of GWs
\begin{equation}
\colboxed{
\OGW(\eta_0,k) = \frac{c_g\Omega_{r,0}}{972} 
  \iint_\S \dd x\,\dd y
  \frac{x^2}{y^2} \left[1-\frac{(1+x^2-y^2)^2}{4x^2} \right]^2
  \Pz(kx) \Pz(ky) 
  \left[ \Ic(x,y)^2 + \Is(x,y)^2 \right].
}
\label{eq: Omega GW with PS}
\end{equation}

%%%%%%%%%%%%%%%%%%%%%%%%%%%%%%%%%%%%%%%%%%%%%%%%%%%%%%%%%%%%
\section{IV. Bispectrum of Gravitational Waves}
\begin{figure}[t!] \centering
$\hbox{ \convertMPtoPDF{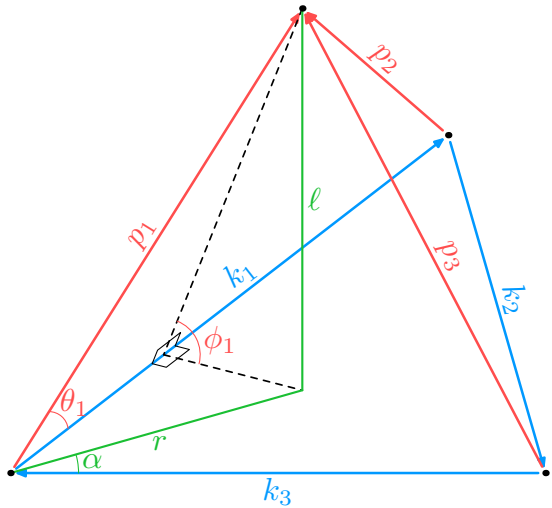}{.8}{.8} }$
\caption{Geometrical configuration for the contraction $(i)$ of the 6-point function of $\zeta$ written in Eq.~\eqref{eq: contraction 3pt (i)}. }
\label{fig: tetrahedron}
\end{figure}

%%%%%%%%%%%%%%%%%%%%%%%%%%%%%%%%%%%%%%%%%%%%%%%%%%%%%%%%%%%%
\noindent
In this section we compute the bispectrum (three-point function) of GWs. 
Let us start from the solution \eqref{eq: h with Ic, Is} for GWs, and write the three-point function as
\begin{multline}
\left \langle h^r(\eta,\vk_1) h^s(\eta,\vk_2) h^t(\eta,\vk_3) \right \rangle = 
  \left(\frac 49 \right)^3 \int \frac{\dd^3 p_1}{(2\pi)^3} \int \frac{\dd^3 p_2}{(2\pi)^3} \int \frac{\dd^3 p_3}{(2\pi)^3} 
  \frac{1}{k_1^3 k_2^3 k_3^3 \eta^3} 
  e^r(\vk_1,\vp_1) e^s(\vk_2,\vp_2) e^t(\vk_3,\vp_3) \cdot \\
  \cdot \Big\langle \zeta(\vp_1) \zeta(\vk_1-\vp_1) \zeta(\vp_2) \zeta(\vk_2-\vp_2) \zeta(\vp_3) \zeta(\vk_3-\vp_3) \Big\rangle 
  \big[ \cos(k_1\eta) \Ic(x_1,y_1) + \sin(k_1\eta) \Is(x_1,y_1)\big] \cdot \\ 
  \cdot \big[ \cos(k_2\eta) \Ic(x_2,y_2) + \sin(k_2\eta) \Is(x_2,y_2)\big]   
  \big[ \cos(k_3\eta) \Ic(x_3,y_3) + \sin(k_3\eta) \Is(x_3,y_3)\big],
\label{eq: 3-pt h def}
\end{multline}
where $x_i=p_i/k_i$ and $y_i=|\vk_i-\vp_i|/k_i$. The details of the 
calculation of the six-point function of the curvature perturbation $\zeta$ are given in Appendix~B. 
We have eight possible contractions for $\vk_i\neq 0$ that yield the same contribution to the bispectrum. We can evaluate the three-point function for any of these configurations and multiply by eight the result. The three-point function (\ref{eq: 3-pt h def}) becomes then (we understand that $\vp_2 = \vp_1-\vk_1$, $\vp_3 = \vp_1+\vk_3$, and $y_1=x_2$, $y_2=x_3$, $y_3=x_1$):
\begin{multline}
\left \langle h^r(\eta,\vk_1) h^s(\eta,\vk_2) h^t(\eta,\vk_3) \right \rangle = 
  (2\pi)^3  \vdelta(\vk_1+\vk_2+\vk_3) 
  \, 8 \left(\frac 49 \right)^3 \pi^3
  \int \dd^3 p_1
  \frac{1}{k_1^3 k_2^3 k_3^3 \eta^3}  \cdot \\
  \cdot e^r(\vk_1,\vp_1) e^s(\vk_2,\vp_2) e^t(\vk_3,\vp_3)
  \frac{\Pz(p_1)}{p_1^3} \frac{\Pz(p_2)}{p_2^3} \frac{\Pz(p_3)}{p_3^3} 
  \left[ \cos(k_1\eta) \Ic\left( \frac{p_1}{k_1},\frac{p_2}{k_1}\right) + \sin(k_1\eta) \Is\left( \frac{p_1}{k_1},\frac{p_2}{k_1}\right)\right] \cdot \\ 
  \cdot \left[ \cos(k_2\eta) \Ic\left( \frac{p_2}{k_2},\frac{p_3}{k_2}\right) + \sin(k_2\eta) \Is\left( \frac{p_2}{k_2},\frac{p_3}{k_2}\right)\right]   
  \left[ \cos(k_3\eta) \Ic\left( \frac{p_3}{k_3},\frac{p_1}{k_3}\right) + \sin(k_3\eta) \Is\left( \frac{p_3}{k_3},\frac{p_1}{k_3}\right)\right].
\label{eq: 3-pt h contracted}
\end{multline}
The polarisation tensors defined in Eq.~\eqref{eq: polarisation tensors spherical} involve the angles $\theta_i$, $\phi_i$ (shown in Fig.~\ref{fig: tetrahedron} for $i=1$) which identify $\vp_i$ in spherical coordinates around the axis $\vk_i$. 
\newline
With reference to Fig.~\ref{fig: tetrahedron}, the vectors $\vk_i$ in blue are given and we can choose a reference frame such that
\begin{equation}
\vk_1 = \left( k_{1x}, k_{1y}, 0 \right) , \quad
\vk_2 = \left( k_{2x}, k_{2y}, 0 \right) , \quad
\vk_3 = \left( -k_{3}, 0, 0 \right);
\label{eq: tetrahedron k_i}
\end{equation}
the quantities $\ell,r$, and $\alpha$ in green are a convenient choice of cylindrical coordinates as integration variables,
\begin{equation}
\int\dd^3p_1 \longrightarrow \int_{-\infty}^{+\infty} \dd \ell \, \int_0^{+\infty} r \,\dd r \int_0^{2\pi} \dd \alpha \,;
\label{eq: tetrahedron int}
\end{equation}
the quantities marked in red give the expressions to plug in Eq.~\eqref{eq: 3-pt h contracted},
\begin{equation}
\begin{gathered}
\vp_1 = \left( r\cos\alpha, r \sin\alpha, \ell \right) , \quad
\vp_2 = \left( -k_{1x} + r\cos\alpha, -k_{1y} + r \sin\alpha, \ell \right) , \quad
\vp_3 = \left( -k_3 + r\cos\alpha, r \sin\alpha, \ell \right) , \\
p_i^2 \sin^2 \theta_i = p_i^2 - \frac{\abs{\vp_i\cdot\vk_i}^2}{k_i^2} , \quad
\sin \phi_i = \frac{\ell\, k_i}{\abs{\vp_i\times\vk_i}} .
\end{gathered}
\label{eq: tetrahedron p_i}
\end{equation}
Eqs.~\eqref{eq: 3-pt h contracted}, \eqref{eq: polarisation tensors spherical},  and \eqref{eq: Ic, Is}, with the replacements listed in~\eqref{eq: tetrahedron k_i}, \eqref{eq: tetrahedron int}, \eqref{eq: tetrahedron p_i}, contain all the ingredients for the numerical calculation of the bispectrum of GWs.

Out of the eight possible polarisations $(r,s,t)$ of the three-point function, four of them vanish due to parity arguments applied to the polarisation tensors, in analogy to what happens for the two-point function, see Eq.~\eqref{eq: polarisation 2-pt parity}.
Among the terms contained in Eq.~\eqref{eq: 3-pt h contracted}, the only ones which are odd under the parity transformation $\ell\to -\ell$ (that is, a parity transformation with respect to the plane containing $\vk_1,\vk_2,\vk_3$) are the polarisation tensors $e^{\times}$, and all other terms are even.
This implies that the only four non-vanishing polarisation combinations for the three-point functions are
\be
(+++), \quad
(+\times \times),\quad
(\times + \times),\quad
(\times \times +).
\label{eq: 3pt nonzero pol}
\ee

%%%%%%%%%%%%%%%%%%%%%%%%%%%%%%%%%%%%%%%%%%%%%%%%%%%%%%%%%%%%
\section{V. Numerical results for the Energy Density and Bispectrum of GWs}
%%%%%%%%%%%%%%%%%%%%%%%%%%%%%%%%%%%%%%%%%%%%%%%%%%%%%%%%%%%%
\label{sec: numerical results}

\subsection{Energy density of GWs}
\noindent
We devote this section to the results of the numerical integration 
for the scalar power spectra $\Pz(k)$ obtained for a few illustrative cases of the mechanism discussed in \cite{Espinosa:2017sgp} and summarised in the Introduction.
We rewrite for convenience the energy density of GWs of Eq.~\eqref{eq: Omega GW with PS} in terms of the variables $(d,s)$ defined in Eq.~\eqref{eq: xy to ds} as
\begin{equation}
\OGW(\eta_0,k) = \frac{c_g\Omega_{r,0}}{36} 
  \int_0^{\frac{1}{\sqrt{3}}}\dd d \int_{\frac{1}{\sqrt{3}}}^{\infty}\hspace{-5pt}\dd s
  \left[ \frac{(d^2-1/3)(s^2-1/3)}{s^2-d^2} \right]^2
  \Pz\left(\frac{k\sqrt{3}}{2}(s+d)\right) \Pz\left(\frac{k\sqrt{3}}{2}(s-d)\right)
  \left[ \Ic(d,s)^2 + \Is(d,s)^2 \right],
\label{eq: Omega GW with PS, ds}
\end{equation}
where the functions $\Ic,\Is$ are defined in \eqref{eq: Ic, Is} and are plotted in Figs. \ref{fig: integral eta 3D} and \ref{fig: integral eta 2D}.

\begin{figure}[h!]
\centering
\includegraphics[width=.7\textwidth]{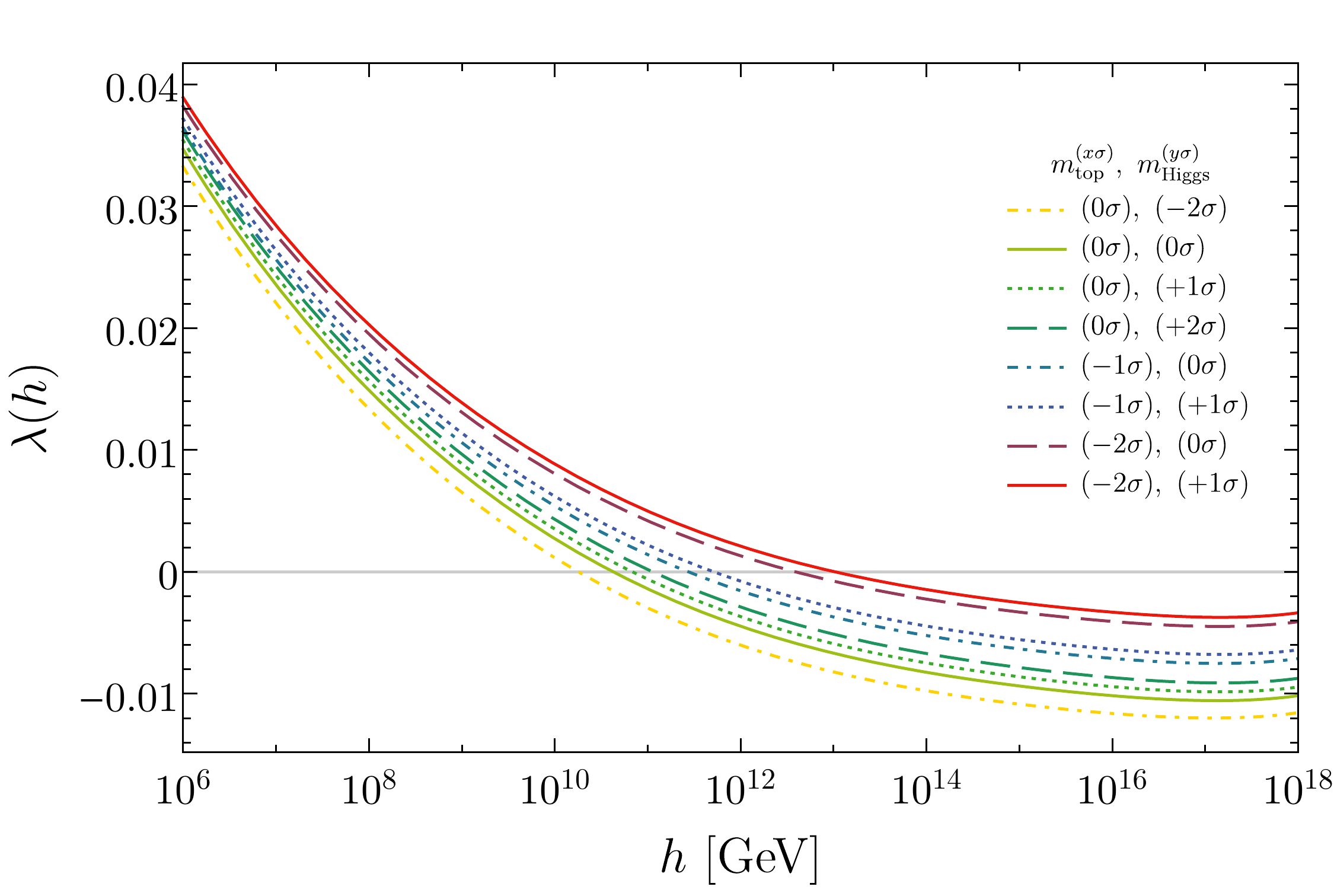}
\caption{Running of the quartic Higgs coupling $\lambda$ for the following Higgs and top masses: 
$m_\text{Higgs}=125.09\pm 0.24$ GeV and $m_{\rm top}=172.47\pm 0.5$  GeV.}
\label{fig: running lambda}
\end{figure}

\begin{figure}[t!]
\centering
\includegraphics[width=.65\textwidth]{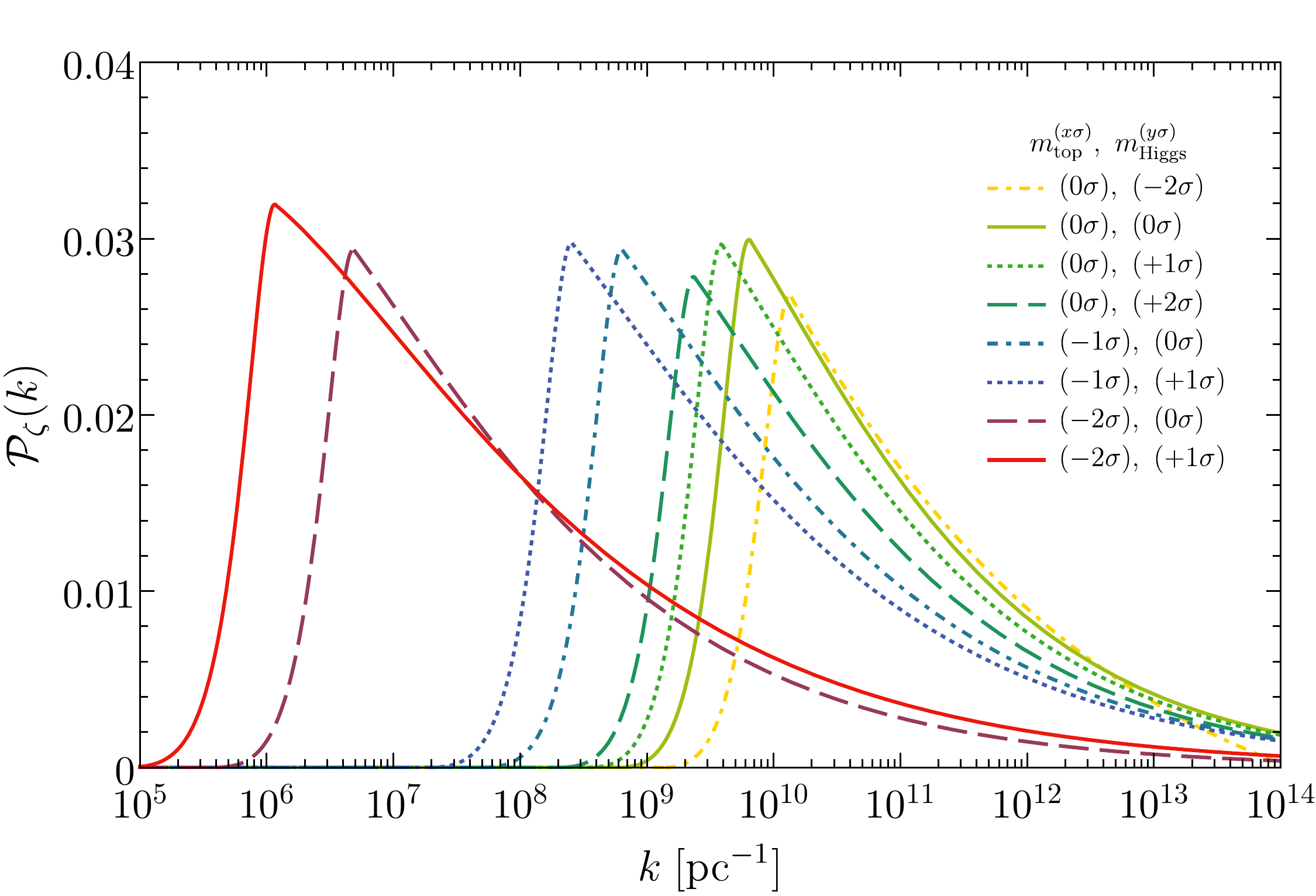}
\caption{The power spectrum of the comoving curvature perturbation during the radiation phase obtained in Ref.~\cite{Espinosa:2017sgp} for the following Higgs and top masses: 
$m_\text{Higgs}=125.09\pm 0.24$ GeV and $m_{\rm top}=172.47\pm 0.5$  GeV.}
\label{fig: PS}
\end{figure}

We consider the running of the quartic Higgs coupling $\lambda$ for some sample points in the parameter space $(m_\text{top},m_\text{Higgs})$ denoted by the number of standard deviations from the measured central values. We have taken the current LHC combination $m_\text{Higgs}=125.09\pm 0.24$ GeV \cite{Mh} and $m_{\rm top}=172.47\pm 0.5$  GeV \cite{Mt}. 
The corresponding running of the quartic Higgs coupling $\lambda$ is shown in Fig.~\ref{fig: running lambda}.
\newline
Each of these points defines therefore a different Higgs potential, for which we run an evolution of the Higgs field completely analogous to what was described in Ref.~\cite{Espinosa:2017sgp}, by keeping a fixed Hubble rate $H=10^{12}$ GeV. 
This evolution leads to the creation of PBH during the radiation dominated era, with a peak in the mass function for scales of the order of $k_*$, the mode that leaves the Hubble radius at the time $t_*$ when the classical evolution of the Higgs field starts, as described in the Introduction.
The corresponding $\Pz$ has basically the same shape in all these cases, and what changes is the reference scale $k_*$ for the enhancement of the power spectrum
%\footnote{The reader might wonder why the power spectrum $P_\zeta$ for the case $m_\text{top}^{(-2.5\sigma)}$ and $m_\text{Higgs}^{(0\sigma)}$ is peaked for modes larger than cases with a less negative running of $\lambda$, contrarily to the trend in  Fig.~\ref{fig: running lambda}. The reason is that we do not change the values of the Hubble rate $H=10^{12}$ GeV and  the reheating temperature, and given that the case $m_\text{top}^{(-2.5\sigma)}$ and $m_\text{Higgs}^{(0\sigma)}$ pushes the instability scale to $\Lambda \lesssim 10^{14}$ GeV, the viable range for the excursion of $h_{\rm c}$ is reduced, and consequently the peak of $\Pz$ is on larger scales. By changing the value of the Hubble rate one can extend the duration of the classical evolution of the Higgs, and consequently shift $\Pz(k)$ to smaller $k$.}
%Appendix C contains the analytical derivation of it and, for the convenience of the reader, 
as we show in Fig.~\ref{fig: PS}.\footnote{If the same mechanism is supposed to give rise to PBHs, then these power spectra yield a final abundance $\Omega_\text{PBH}/\Omega_\text{CDM}$ ranging between $10^{-3}$ and $10^{-1}$ when no accretion is included \cite{Espinosa:2017sgp}.}
\begin{figure}[t!]
\centering
\includegraphics[width=.65\textwidth]{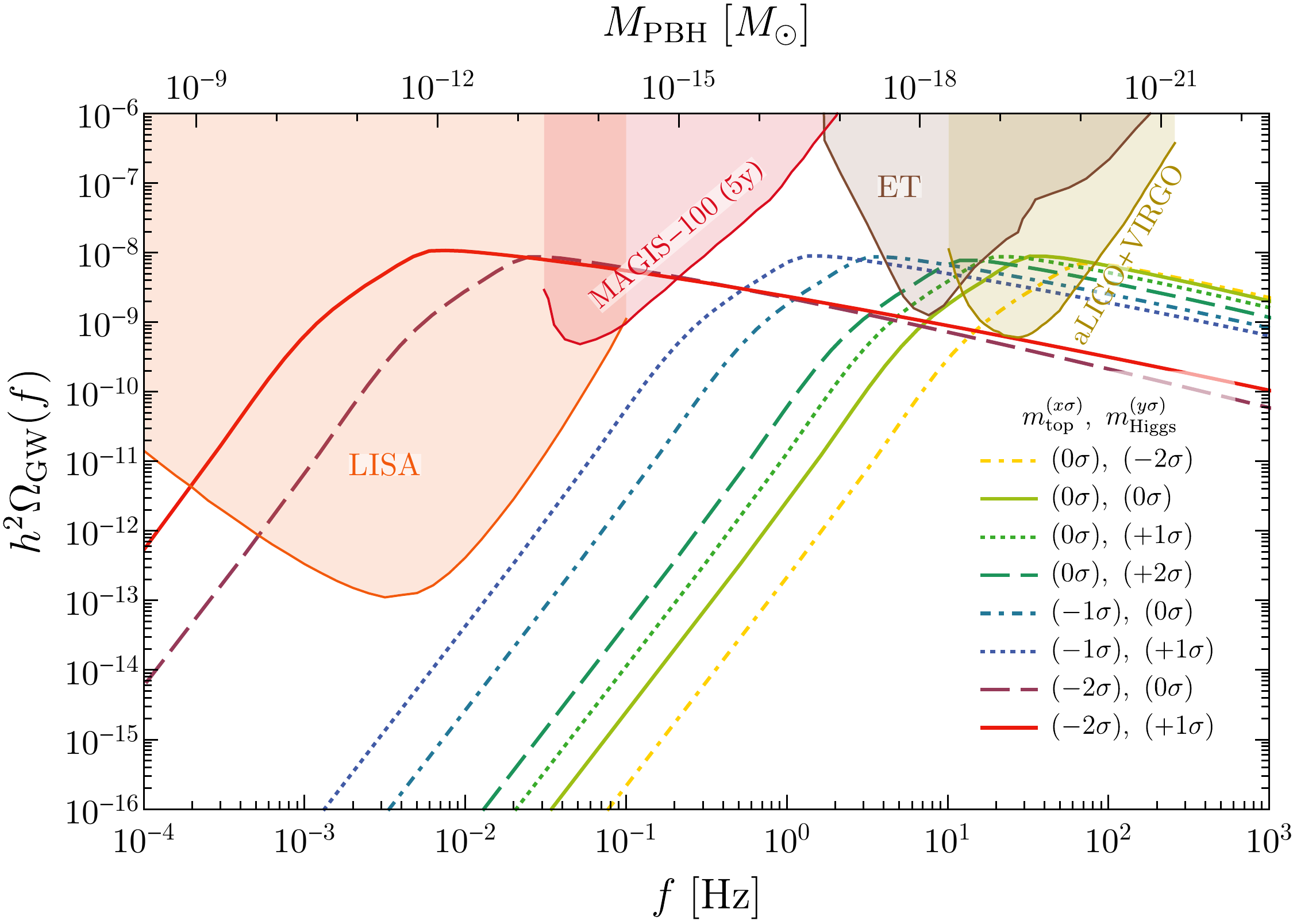}
\caption{Power spectra of GWs for the scalar power spectra generated by the mechanism discussed in Ref. \cite{Espinosa:2017sgp}, compared with the estimated sensitivities for LISA, the Einstein Telescope, MAGIS-100, and the design sensitivity of Advanced LIGO + Virgo. The Higgs and top mass values are $m_\text{Higgs}=125.09\pm 0.24$ GeV and $m_{\rm top}=172.47\pm 0.5$ GeV.}
\label{fig: OmegaGW}
\end{figure}
\newline
The final result for the power spectra of GW is shown in Fig.~\ref{fig: OmegaGW}, together with the comparison with the projected sensitivity of proposed future experiments.
The sensitivity curve for LISA is estimated on the basis of the proposal \cite{Audley:2017drz}: the proposed design (4y, 2.5 Gm of length, 6 links) is expected to yield a sensitivity in between the ones dubbed C1 and C2 in  Ref. \cite{Caprini:2015zlo}\footnote{We thank G. Nardini for clarifying discussions about this point.}.
We also include the projected design sensitivity for Advanced LIGO + Virgo from Ref. \cite{TheLIGOScientific:2016dpb}, the estimated sensitivity for the proposed Einstein Telescope (ET) \cite{ET-2}, and the estimated reach for the 5-years program of MAGIS-100 at FERMILAB \cite{MAGIS}.

The GW power spectra are shown for different combinations of the values of the Higgs and top masses where the symbols
$m_\text{Higgs}^{(\pm n\sigma)}$ and $m_{\rm top}^{(\pm n\sigma)}$ indicate their values $\pm n\sigma$ away from their central values. 
A GW power spectrum for values of the Higgs boson mass  $m_\text{Higgs}=125.09$ GeV (the current central value) and $m_{\rm top}=171.47$  GeV, is well within the reach of LISA. To relate the amount of GWs and the PBH abundance at formation following the proposal in Ref.~\cite{Espinosa:2017sgp}, one can use the relation $M_{\rm PBH}\simeq 50 M_\odot(10^{-9}{\rm Hz}/f)^2$. In Fig.~\ref{fig: OmegaGW}, we have used that  relation to translate the frequencies of the GW signal in terms of the  peak mass of the PBH distribution.

One fundamental information to be drawn  from Fig. \ref{fig: OmegaGW} is that the frequency at the peak depends in a sensitive way on the Higgs and top masses, ranging from $10^{-2}$ to about 10 Hz, see Table I. 
Therefore, according to the Higgs and top masses, the signal falls either within the LISA or the ET and Advanced-Ligo sensitivity
curves. 
This implies that a detected signal can be cross-checked with the information obtained through colliders, thus either confirming or ruling out its Standard Model origin. 

We draw the attention of the reader that our results for the GW power spectra in Fig.  \ref{fig: OmegaGW} are sensitive to the value of the Higgs field at the beginning of its classical dynamics. A per mill
change in such a value can lead to  variations of the power spectrum of the curvature perturbation by $(2-4)$ orders of magnitude. However, from Fig. \ref{fig: OmegaGW} it is clear that we can still afford a change in ${\cal P}_\zeta$ of three orders of magnitude.

\subsection{The spectral tilt of GWs at low and high frequencies}
\noindent
As we have mentioned in the introduction, the spectral tilt of the  GW spectrum is a very interesting observable  as GWs cover a large range of frequencies. 
For instance, writing the  GW energy density as $\Omega_{\rm GW}=\Omega_{\rm GW}^{\rm CMB}(f/f_{\rm CMB})^{n_T}$, being $n_T$ the spectral tilt and $f_{\rm CMB}\sim 7.7\cdot 10^{-17}$ Hz the CMB frequency, a limit of $n_T\lsim 0.35$ can in principle  be obtained for the best LISA configuration with six links, five million km arm length and a five year mission \cite{LISA}.

If the scalar power spectrum $\Pz(k)$ is vanishing or negligible for $k$ smaller than some scale $k_*$, and approximately constant for $k>k_*$ as in our case, then at small $k$ we have $\OGW\sim k^3$. Indeed, in this case $\Pz(kx)$ in Eq.~\eqref{eq: Omega GW with PS, ds} for $k\ll k_*$ selects $s\gtrsim 1/k$ in the integral over $s$, so that  the tail at high $s$ of $\Ics(d,s)$ is peaked up and it  goes as $1/s^2$ (see Fig.~\ref{fig: integral eta 2D}). The resulting   overall integral is therefore of order 
\be
 \int_{1/k}\frac{\dd s}{s^{4}}\sim k^3.
\ee
As for the spectral  tilt  at $k\gg k_*$, the integral over $s$ in Eq.~\eqref{eq: Omega GW with PS, ds} is peaked at $s\sim \sqrt 3$ due to the spike in $\Ics(d,s)$ (see Fig.~\ref{fig: integral eta 2D}) and the dependence on $k$ comes from $\Pz(k x) \Pz(k (\sqrt 3-x))$ which has a spectral tilt equal roughly to twice the spectral index of $\Pz$. In our case, $\Pz(k)\sim k^{-0.35}$ and $\OGW(k)$ turns out to go as $\sim k^{-0.6}$. 
For a narrow scalar power spectrum $\Pz(k)$, we would expect by similar arguments a spectral index $\sim +2$ at small $k$ and a quite sharp cutoff at high $k$.

The final parametrisation of the GW spectrum induced by the Higgs fluctuations is therefore
\be
\boxed{
\Omega_{\rm GW}(f)\simeq 3\cdot 10^{-8}\left(\frac{f}{f_*}\right)^{n_T}\,\,\,\, {\rm with}\,\,\,\,n_T= \left\{ 
\begin{array}{ll} 
3 \ \ & {\rm for}\ f<f_*, \\ 
-0.6 \ \ & {\rm for}\ f>f_*.
\end{array} 
\right..}
\label{eq: OmegaGW nT}
\ee
%with 
%
%\be
%\boxed{
%n_T= \left\{ 
%\begin{array}{ll} 
%3 \ \ & {\rm for}\ f<f_*, \\ 
%-0.6 \ \ & {\rm for}\ f>f_*.
%\end{array} 
%\right.
%}
%\ee
\begin{table}[h!]
\begin{tabular}{ccc}
\toprule
$m_\textrm{top}^{(x\sigma)}$&$\ m_\textrm{Higgs}^{(y\sigma)}$ & $f_*({\rm Hz})$ \\
\midrule
$(0\sigma)$&$(-2\sigma)$ &   40.0 \\$(0\sigma)$&$(0\sigma)$ &   21.8 \\
$(0\sigma)$&$(+1\sigma)$ &  13.0 \\
$(0\sigma)$&$(+2\sigma)$ &  7.74 \\
$(-1\sigma)$&$(0\sigma)$ &  2.02 \\
$(-1\sigma)$&$(+1\sigma) $& 0.80 \\
%$(-2.5\sigma)$&$(0\sigma)$ & 0.18 \\
$(-2\sigma)$&$(0\sigma)$ &  0.015 \\
$(-2\sigma)$&$(+1\sigma)$ & 0.0038 \\
\bottomrule
\end{tabular}
\caption{Values of $f_*$ defined in Eq.~\eqref{eq: OmegaGW nT} for each of the cases considered in Fig.~\ref{fig: OmegaGW}.}
\label{tab: f*}
\end{table}
\newline
The values of $f_*$ for the cases we consider are listed in Table~\ref{tab: f*}.
The parametrisation of Eq.~\eqref{eq: OmegaGW nT} is useful to deduce its  detectability by LISA.  The investigation of a generic GW background whose energy density is parametrised as $\Omega_{\rm GW}(f)=A(f/f_*)^{n_T}$ can be found in Ref.~\cite{LISA} where it was imposed that the  signal-to-noise ratio is  larger than 10, see Fig.~2 of Ref.~\cite{LISA}. It seems that for the case of  Higgs mass  $m_\text{Higgs}=125.09$ GeV and $m_{\rm top}=171.47$  GeV not only  the  amplitude of the gravitational waves from Higgs perturbations, but also its spectral index can be measured with accuracy, opening the possibility of a full identification of the underlying mechanism.

 Were GWs found, the value of the frequency $f_*$ would allow to identify the approximate position of the instability scale $\Lambda_I$ of the Higgs potential, defined by $V(\Lambda_I)=0$. 
%We show in Fig.~\ref{fig: fit f* Lambda} the results we obtained for the cases under consideration, together with a fit with a power-law relation. 
The instability scale $\Lambda_I$ can be identified by the relation
\begin{equation}
\boxed{
\Lambda_I \simeq 3 \cdot 10^{11} \left(\frac{f_*}{\text{Hz}} \right)^{-0.65} \text{ GeV}.}
\end{equation}
We stress that this relation is robust in the sense that the frequency changes very little even when the overall amplitude of the GW signal decreases due to a variation of the initial condition of the classical
Higgs field.
It is remarkable that a measurement of the frequency of the GW signal can be directly related to such a high energy scale.

%\begin{figure}[h]
%\includegraphics[width=.65\textwidth]{Lambda-fstar}
%\caption{Central frequency $f_*$ for the GW signal plotted against the instability scale $\Lambda$ for each of the couples of masses $(m_\text{top},m_\text{Higgs})$ of Table~\ref{tab: f*}, together with an approximate fit.}
%\label{fig: fit f* Lambda}
%\end{figure}

\subsection{The three-point correlator of GWs and its consistency  relations}
\noindent
In this subsection we present our findings for the three-point correlator of the GWs. 
As mentioned in the introduction, the community has already started discussing the   detectability of such  non-Gaussian signal  at interferometers \cite{three}. 
The ultimate reason for measuring the GW bispectrum is   to exploit  the correspondence between the  three-point and the two-point correlators in order to discriminate the different mechanisms which give rise to a GW signal hopefully measured by LISA. 
Unfortunately, such a non-Gaussian signal seems not an observable quantity with a GW detector, which unavoidably detects the sum of the GW signals from all patches of the sky. The sum of all these signals, independently from their primordial correlation, results as a Gaussian signal by virtue of the Central Limit Theorem (see \cite{Bartolo:2018rku} for a detailed discussion).

We define the bispectrum $B_h^{rst}(\vk_1,\vk_2,\vk_3)$  (the temporal dependence on $\eta$ is understood) as
\begin{equation}
\left \langle h^r(\eta,\vk_1) h^s(\eta,\vk_2) h^t(\eta,\vk_3) \right \rangle \equiv 
  (2\pi)^3  \vdelta(\vk_1+\vk_2+\vk_3)
  \, B_h^{rst}(\vk_1,\vk_2,\vk_3).
\label{eq: bispectrum}
\end{equation}
We also define a dimensionless normalised shape $\Bh^{rst}(\vk_1,\vk_2,\vk_3)$ in order to cancel the time scaling of GWs as $1/\eta$ [see Eq.~\eqref{eq: h with Ic, Is}]: 
\begin{equation}
\Bh^{rst}(\vk_1,\vk_2,\vk_3) = k_1^{2} k_2^{2} k_3^{2} \frac{B_h^{rst}(\vk_1,\vk_2,\vk_3)}{\sqrt{\Ph(k_1)\Ph(k_2)\Ph(k_3)}} \,,
\label{eq: bispectrum normalised}
\end{equation}
where $\Ph(k)$ is the dimensionless power spectrum defined in Eq.~\eqref{eq: def P h}.
As for the oscillatory behaviour of the two- and three-point functions, we consider their envelope in time.
This simplification is not physically adequate when considering the three-point function measured by a GW detector, given that the overall phase of would depend on the direction of propagation of the incoming wave and is crucial when assessing the non-Gaussianity of the signal. 
In the present discussion, it is motivated by our interest in the amplitude of the primordial bispectrum which could be measurable in principle on a constant time hypersurface.
We replace then the oscillating function in squared brackets in the solution~\eqref{eq: h with Ic, Is} by its envelope
\begin{equation}
\Ic(x,y) \cos(k\eta) + \Is(x,y) \sin(k\eta) \to \sqrt{\Ic(x,y)^2 + \Is(x,y)^2} \,,
\end{equation}
both for $B_h$ and $\Ph$ in Eq.~\eqref{eq: bispectrum normalised}.

\begin{figure}[h]
\includegraphics[width=.49\textwidth]{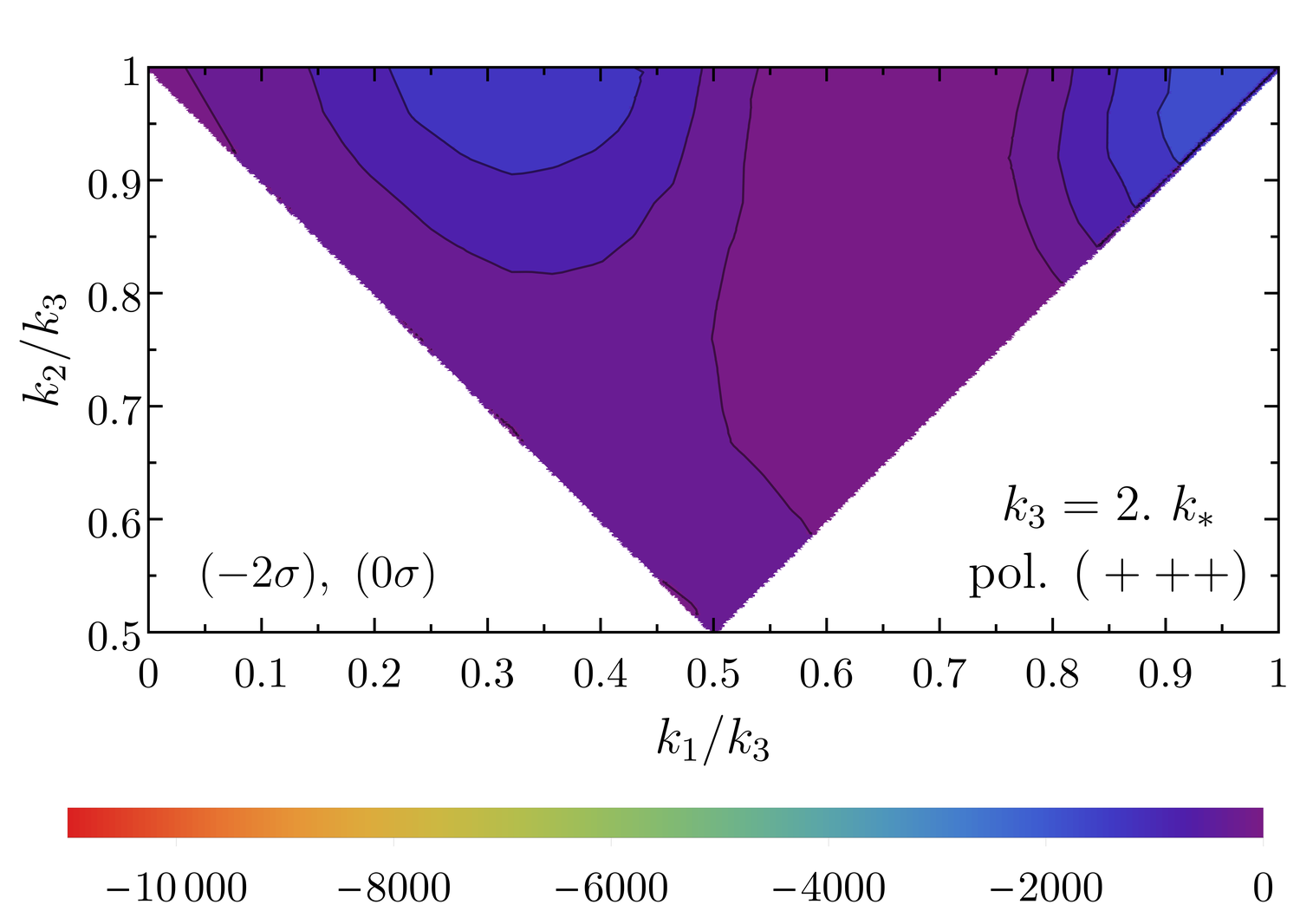}
\includegraphics[width=.49\textwidth]{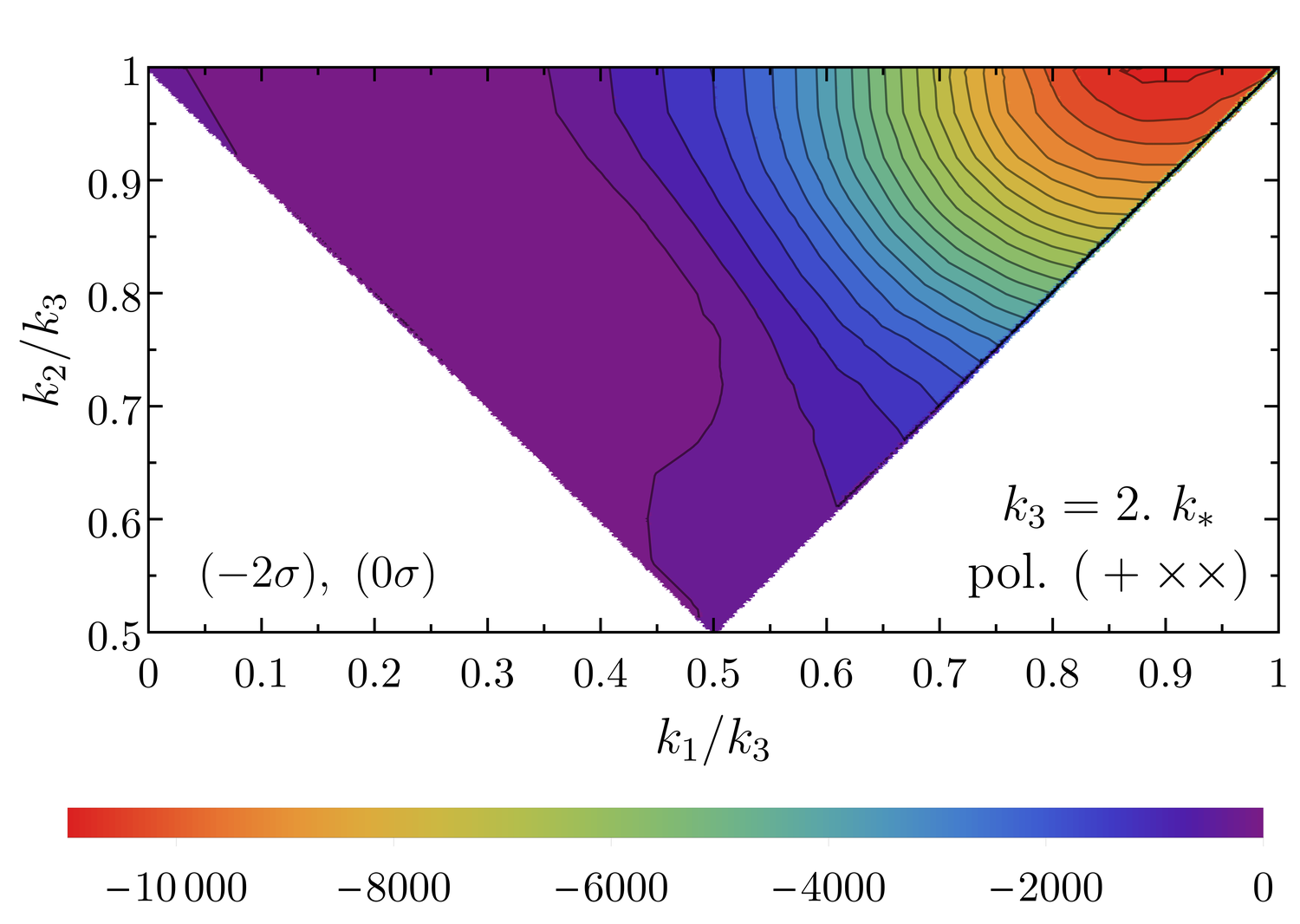} \newline
\includegraphics[width=.49\textwidth]{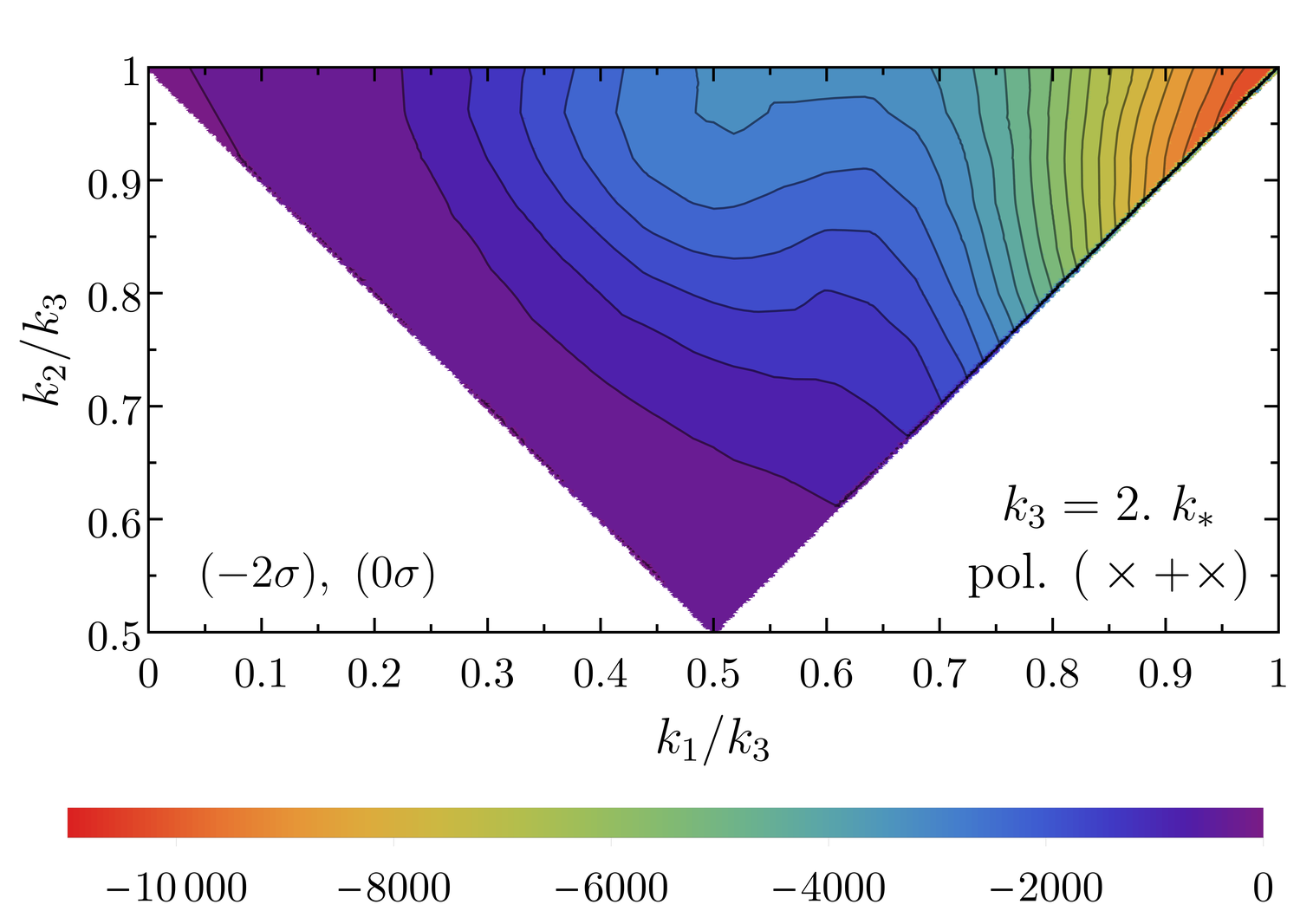}
\includegraphics[width=.49\textwidth]{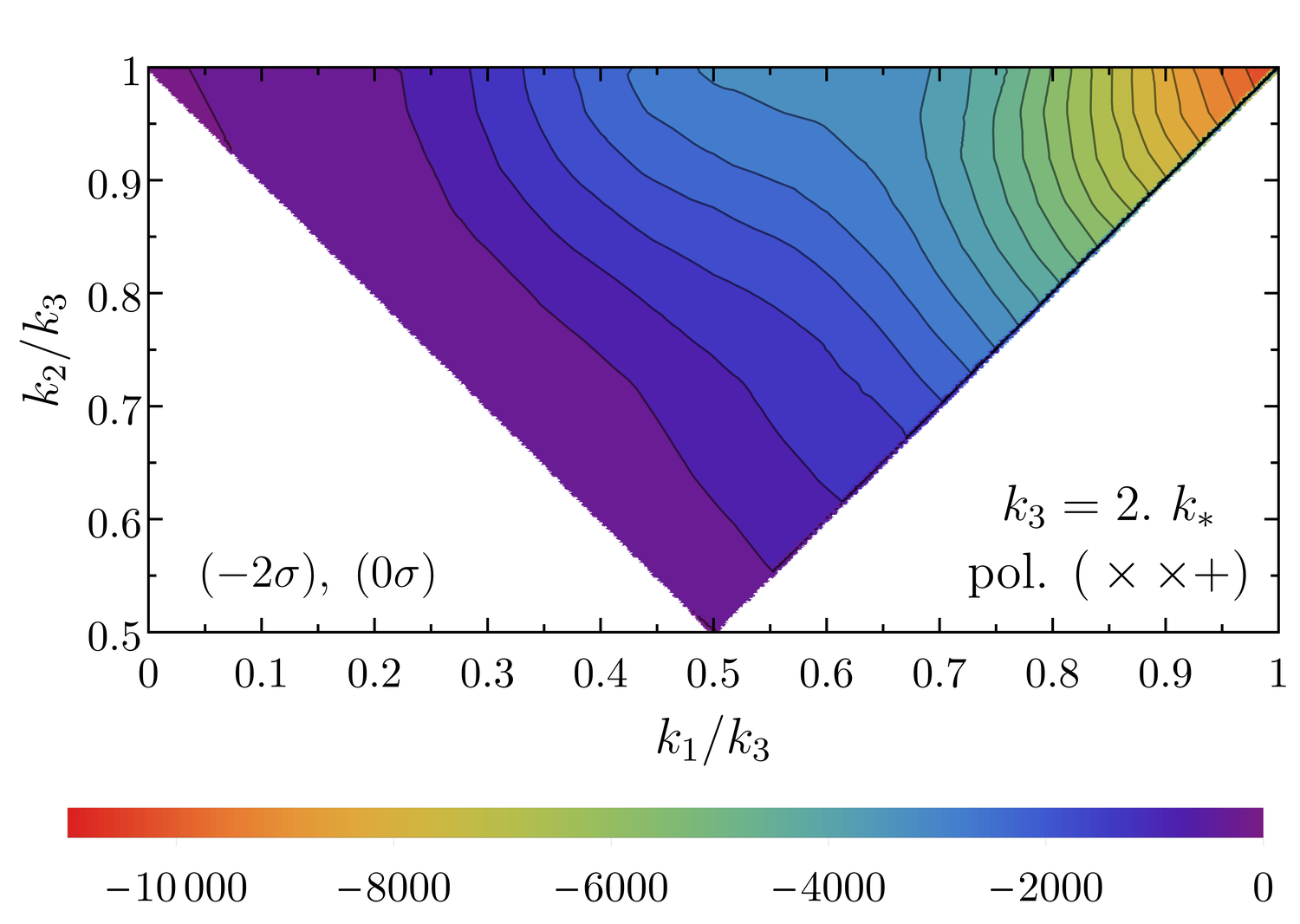} \newline
\includegraphics[width=.49\textwidth]{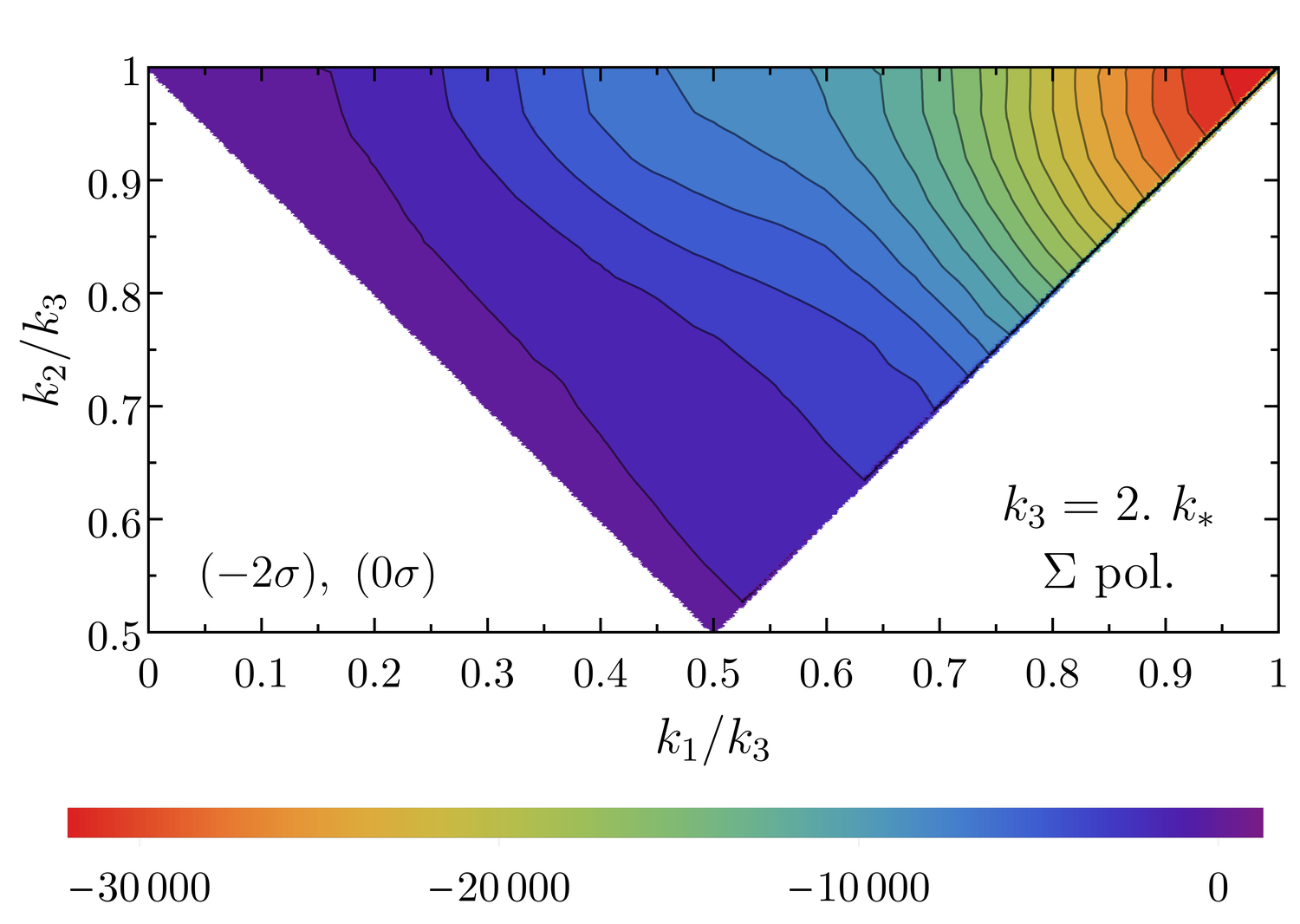}
\includegraphics[width=.49\textwidth]{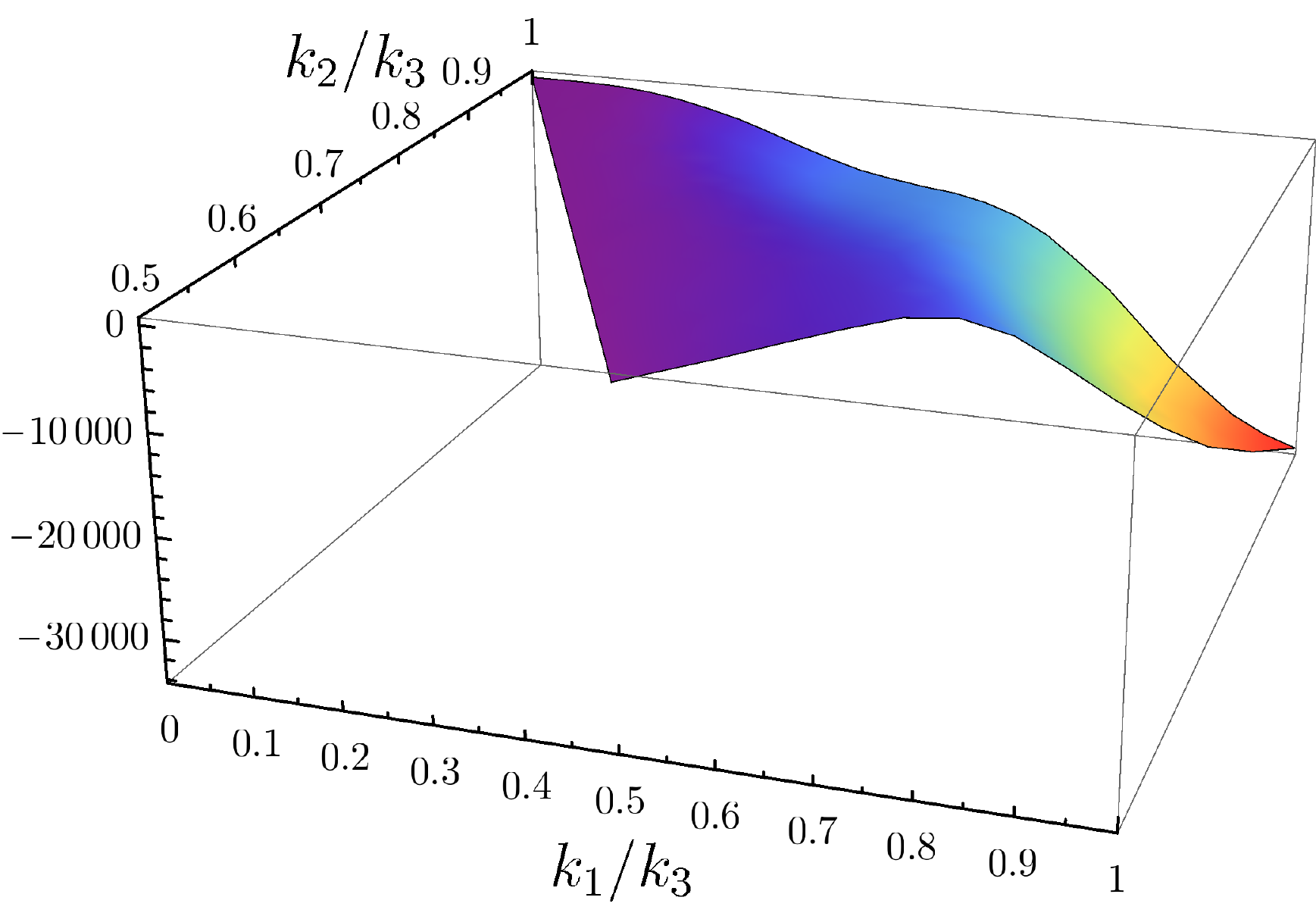}
\caption{Normalised shapes of GWs [defined in Eq.~\eqref{eq: bispectrum normalised}] for the spectrum in the case $(m_\textrm{top}^{(-2\sigma)},\ m_\textrm{higgs}^{(0\sigma)})$. Here $k_3$ is fixed to be $2k_* $, corresponding to $0.04$ Hz. 
The upper four plots show the four non-vanishing polarisations listed in Eq. (\ref{eq: 3pt nonzero pol}). The two plots at the bottom show the sum over all the polarisations.}
\label{fig: bispectra GW k2}
\end{figure}

\begin{figure}[h!]
\includegraphics[width=.49\textwidth]{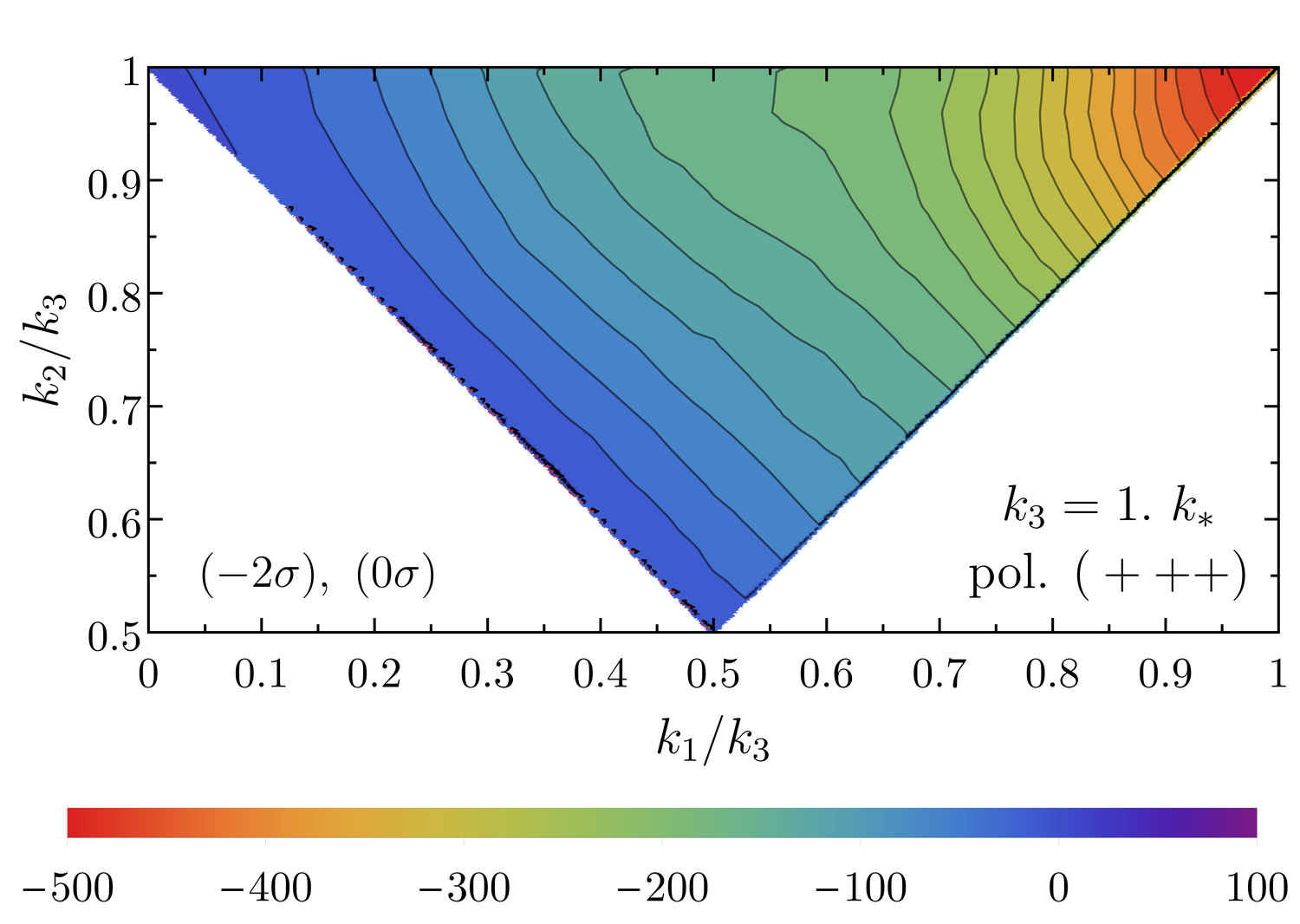}
\includegraphics[width=.49\textwidth]{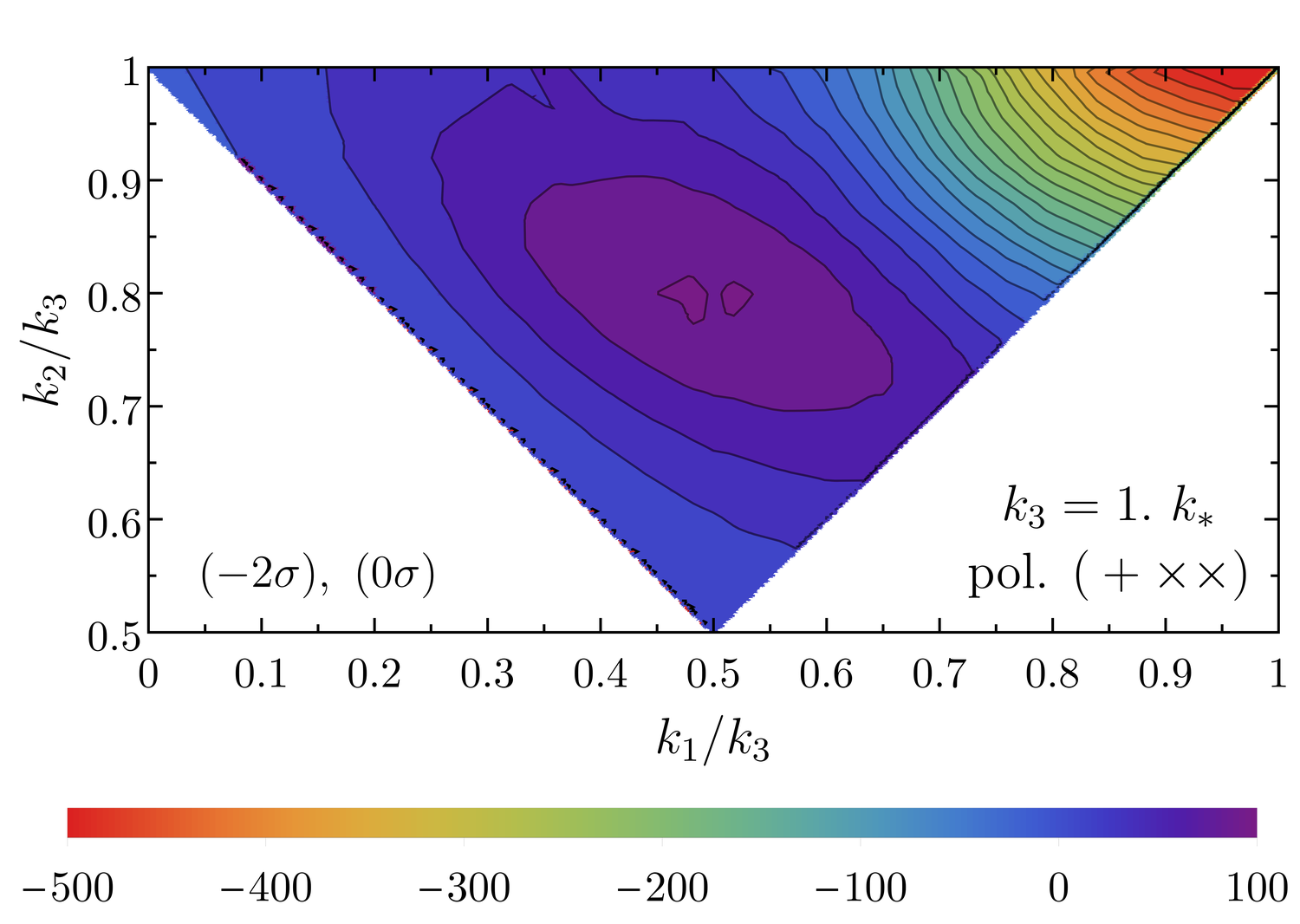} \newline
\includegraphics[width=.49\textwidth]{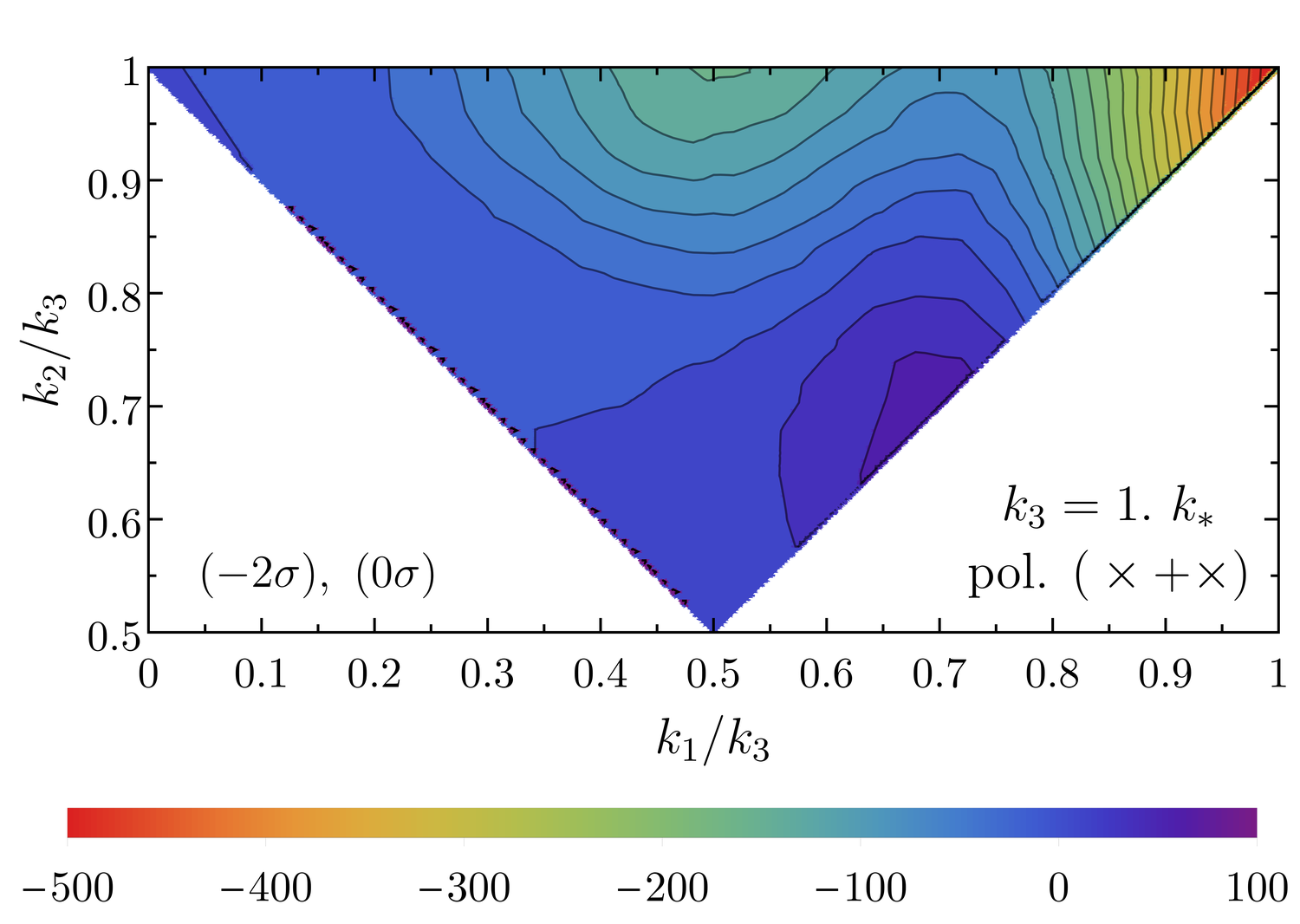}
\includegraphics[width=.49\textwidth]{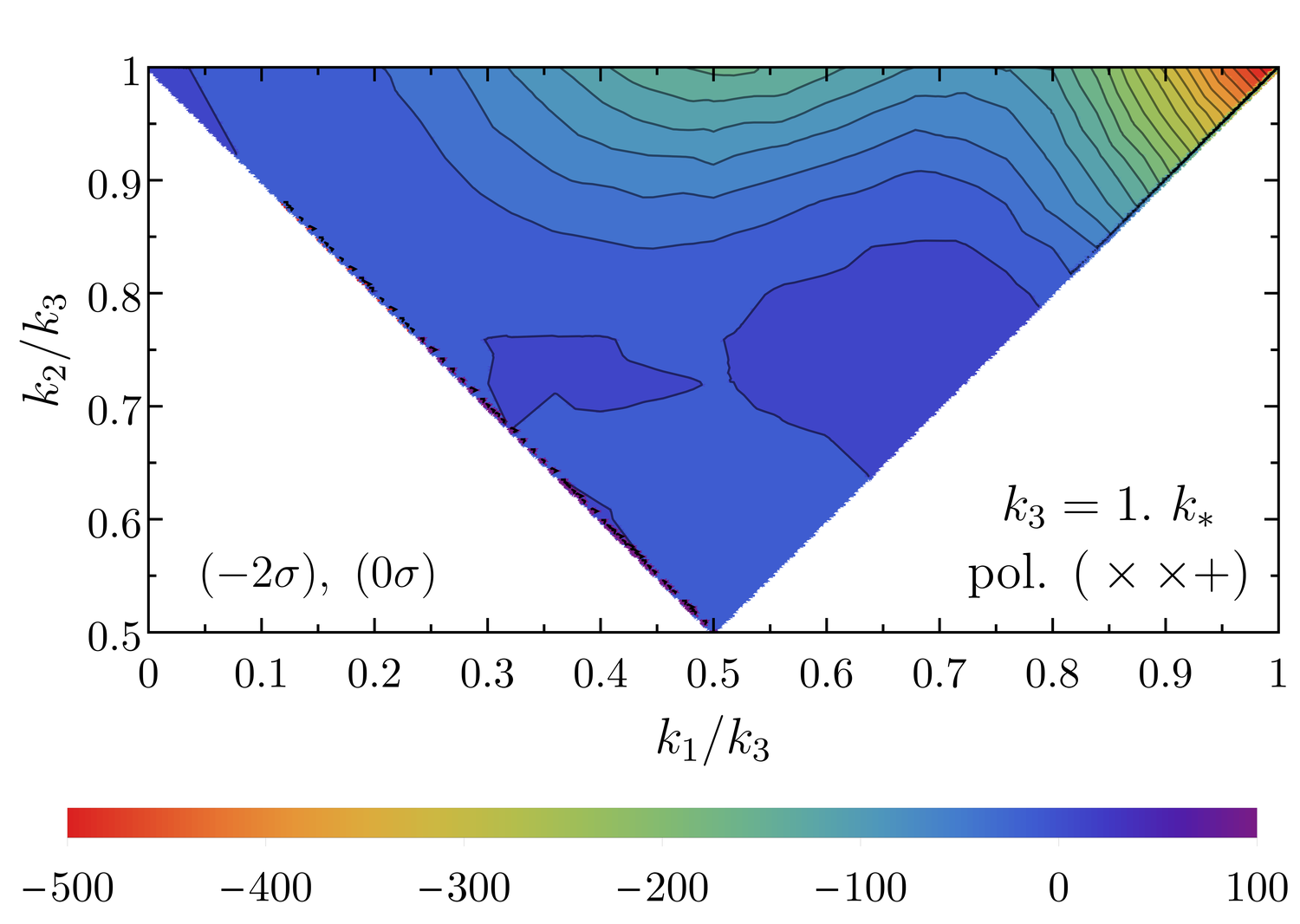} 
\caption{Same as Fig.~\ref{fig: bispectra GW k2}, for  $k_3=k_* $, corresponding to $0.02$ Hz.}
\label{fig: bispectra GW k1}
\end{figure}

We show the numerical results for the bispectrum by fixing the value of $k_3$ and by ordering the momenta as $k_1\leq k_2 \leq k_3$. 
Figs.~\ref{fig: bispectra GW k2} and \ref{fig: bispectra GW k1} show contours of $\Bh(\vk_1,\vk_2,\vk_3)$ in the plane $(k_1/k_3,k_2/k_3)$ for two values of $k_3$ close to the maximum of $\Pz(k)$ (shown in Fig.~\ref{fig: PS}). 
We choose the case $(m_\textrm{top}^{(-2\sigma)},\ m_\textrm{higgs}^{(0\sigma)})$, as it falls into the window detectable by LISA, but we notice that the result is identical for the other cases, given that the shape of the power spectrum is identical, up to a rescaling of the momenta $(k_1,k_2,k_3)$.
We also notice that the normalised shape defined  in Eq.~\eqref{eq: bispectrum normalised} is invariant under rescaling of the scalar power spectrum $\Pz(k)$. 
In Fig.~\ref{fig: bispectra GW k2}, together with the separate plots for each polarisation, we also show their sum in the lower two plots, both with contours and with a three-dimensional plot.
\newline
From these numerical results we observe several features. First of all, we remind the reader that  there are traditionally several configurations one can analyse: the local one where the signal is peaked for squeezed configurations $k_1\ll  k_2 \simeq  k_3$; the equilateral  configuration peaks for equilateral configurations $k_1\simeq k_2\simeq k_3$ for which the strongest correlations between fluctuation modes  happen when  they cross the horizon approximately at the same time; the  folded configuration for which the signal is boosted for  $k_1 + k_2 \simeq k_3$; and finally the  orthogonal configuration ($k_1\simeq k_2$) which creates a signal with a positive peak at the equilateral configuration and a negative peak at the folded configuration. 

The signal is peaked in the equilateral configuration. 
This does not come as a surprise as the GWs are generated at Hubble crossing and the source depends on spatial gradients of the comoving curvature perturbations and this tends to enhance the signal when the scales involved are not too different.
As a rule of thumb we can propose the following consistency relation for the largest signals
\be
\boxed{
\begin{aligned}
\Bh^{+++}&={\cal O}(-10^3) \,\,\,\,\textrm{for equilateral configurations},\\
\Bh^{+\times\times}&={\cal O}(-10^4) \,\,\,\,\textrm{for equilateral configurations}.
\end{aligned}
}
\ee
As for the signal summed for all the polarizations, the results are presented in the lower plot of Fig.~\ref{fig: bispectra GW k2}. 
From it we can estimate
\be
\boxed{
\begin{aligned}
 \sum_{\rm pol}\Bh&={\cal O}(-3\cdot 10^4) \,\,\,\,\textrm{for equilateral configurations}.
\end{aligned}
}
\ee
%\clearpage

%%%%%%%%%%%%%%%%%%%%%%%%%%%%%%%%%%%%%%%%%%%%%%%%%%%%%%%%%%%%
\section{VI. Conclusions}
%%%%%%%%%%%%%%%%%%%%%%%%%%%%%%%%%%%%%%%%%%%%%%%%%%%%%%%%%%%%
\noindent
In this paper we have characterized the GW signal possibly originated by physics of the Standard Model and its inherent instability scale  appearing in the Higgs scalar sector.
In this sense,  GW physics   allows a test, albeit indirect, of the behaviour of the Standard Model  at large field values. The source of the GWs is generated by the Higgs perturbations created during a primordial epoch of inflation and amplified during the phase in which the Higgs probes the unstable part of the potential.

The energy density $\Omega_{\rm GW}$ can be as large as  $10^{-8}$ and therefore measurable either by LISA or by the ET and Advanced-Ligo, the amplitude being  sensitive to  the initial conditions of the Higgs classical dynamics. 
Which experiment turns out to be relevant is dictated by the frequency at the peak of the signal, which in turn depends on the Higgs and top masses. This is indeed a bonus. The more knowledge
from collider physics is collected on these masses, the more one could confirm or disprove the hypothesis that these GWs come from Standard Model physics.

We have also  characterized the signal in terms of its spectral index as well as three-point correlator.  
This non-Gaussian feature of the signal would not be observable unfortunately in a GW detector, which could measure only the sum of signals from many patches in the sky, whose phases would have been further decorrelated by the propagation in the inhomogeneous background. 
Therefore a detection of both a two- and three-point function of a signal would indicate its non-primordial origin.

We close with some comments. The mechanism described in this paper makes use of the fact that we identify our observed Universe as one of those regions which have been thermally
saved during the reheating stage following  inflation after the Higgs has probed the unstable part of its potential during inflation. 
The choice of the parameters might therefore seem fine-tuned. However, anthropic arguments  come to the rescue as the very same dynamics might create  the dark matter of the Universe under the form of PBHs \cite{Espinosa:2017sgp}. 
Put in other words, if the dark matter has to be ascribed to the Standard Model, then one should also detect the corresponding GW signal.

\bigskip
\centerline{\bf Note added}
\smallskip
\noindent
In this paper we have calculated the amount of gravitational waves induced by the Higgs perturbations generated thanks to the Higgs vacuum instability. This result is independent from the possibility that  the same Higgs perturbations are responsible for (a fraction of) the dark matter in the universe. Even assuming that the latter comes from physics beyond the Standard Model, the gravitational waves can be a cosmological signature of the Higgs vacuum instability.
More comments on the PBH issue and fine-tuning can be found in Ref.~\cite{Espinosa:2018euj}.

\bigskip
\centerline{\bf Acknowledgements}
\smallskip
\noindent
We thank Nicola Bartolo and Germano Nardini for comments and for many useful discussions on LISA. 
We thank also Kazunori Kohri and Takahiro Terada for useful discussions about the comparison of their results, and Subodh Patil for useful correspondence.
A.R.\ and D.R.\ are supported by the Swiss National Science Foundation (SNSF), project {\sl Investigating the Nature of Dark Matter}, project number: 200020-159223.
The work of J.R.E. has been partly supported by the ERC
grant 669668 -- NEO-NAT -- ERC-AdG-2014, the Spanish Ministry MINECO under grants  2016-78022-P and
FPA2014-55613-P, the Severo Ochoa excellence program of MINECO (grants SEV-2016-0588 and SEV-2016-0597) and by the Generalitat de Catalunya grant 2014-SGR-1450.

\appendix
\begin{appendices}

%%%%%%%%%%%%%%%%%%%%%%%%%%%%%%%%
\section{Appendix A:  Dynamics of the Higgs hitting the pole}
%%%%%%%%%%%%%%%%%%%%%%%%%%%%%%%%
\label{sec: app Higgs pole}
\setcounter{equation}{0}
\renewcommand{\theequation}{A.\arabic{equation}}

\noindent
To understand Eq. (\ref{yx}) one solves the equation
\be
\ddot{h}_{\rm c}-\lambda h_{\rm c}^3=0,
\ee
Taking the initial conditions $ h_{\rm c}(0)=h_0$ and $\dot{ h}_{\rm c}(0)=\dot h_0$, and using the fact that there is an integral of motion
\be
\frac{1}{2}\dot{h}_{\rm c}^2-\frac{\lambda}{4}h_{\rm c}^4=-E=\frac{1}{2}\dot{h}_0^2-\frac{\lambda}{4}h_0^4,
\ee
one finds the solution 
\be
 h_{\rm c}(t)=h_0\alpha_0\,{\rm cn}\left(i\sqrt{\lambda}h_0\alpha_0\,t+{\rm cn}^{-1}(1/\alpha_0,1/2), 1/2\right),
\ee
where ${\rm cn}(z,k)$ is one of the Jacobian elliptic functions
and
\be
\alpha_0\equiv \left(1-\frac{2\dot{h}_0^2}{\lambda h_0^4}\right)^{1/4}.
\ee
%Using the identity ${\rm cn}(i z,k)= {\rm nc}(z,k')$, where $k'=\sqrt{1-k^2}$, we find 
%\be
% h_{\rm c}(t)=h_0\,{\rm nc}\left(\sqrt{\lambda}h_0\,t, \frac{\sqrt{3}}{2}\right),
%\ee
The function ${\rm cn}( ix,1/2)$ has poles at $x= K(1/2)$ with residue $-i\sqrt{2}$, where 
\begin{equation}
K(k)=\int_0^{\pi/2}\,\frac{{\rm d}\theta}{\sqrt{1-k\sin^2\theta}}.
%\nonumber\\
%K'(k)&=&K(k'),\nonumber\\
%k'&=&\sqrt{1-k^2}.
\end{equation}
Around the pole the classical value of the Higgs can therefore be approximated by Eq.~\eqref{yx} with
\be
t_{\rm p}=\frac{1}{\sqrt{\lambda}h_0\alpha_0}
\left[K(1/2)+i\,{\rm cn}^{-1}(1/\alpha_0,1/2)\right].
\ee

\setcounter{equation}{0}
%%%%%%%%%%%%%%%%%%%%%%%%%%%%%%%
\section{Appendix B: Four and Six-Point Funtions of the Curvature Perturbation}
%%%%%%%%%%%%%%%%%%%%%%%%%%%%%%%
\renewcommand{\theequation}{B.\arabic{equation}}
\subsection{Four-point function of the curvature perturbation}
\begin{figure}[h!] \centering
\raisebox{-\height}{$\hbox{ \convertMPtoPDF{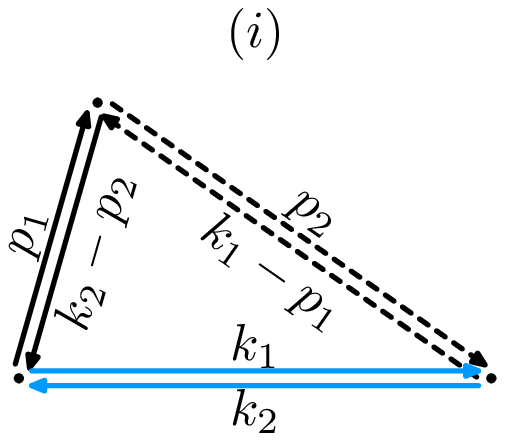}{0.7}{0.7} }$}
\raisebox{-\height}{$\hbox{ \convertMPtoPDF{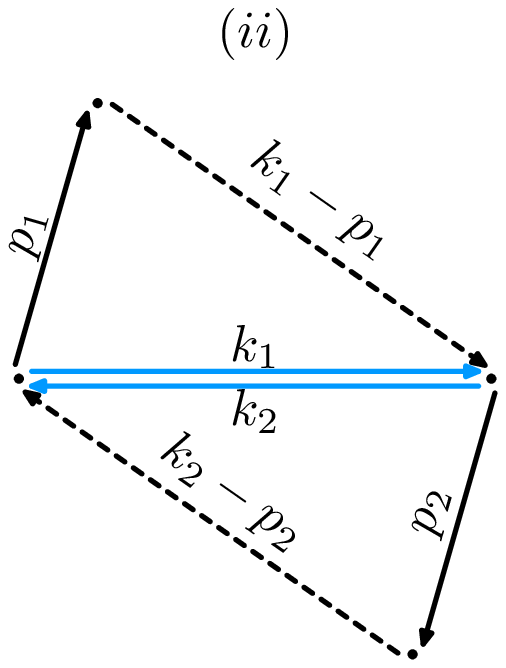}{0.7}{0.7} }$}
\caption{Geometrical configurations for the non-vanishing contractions of the two-point function listed in Eq.~\eqref{eq: contractions two-point}. }
\label{fig: contractions two-point}
\end{figure}

\noindent
The four-point function of the curvature perturbation $\zeta$ in the first line of \eqref{eq: two-pt h def} has two possible non-vanishing contractions for $\vk_1, \vk_2\neq 0$
\begin{equation}
\begin{aligned}
(i)\qquad  & \bcontraction
  { \Big\langle \zeta(\vp_1) }
  { \zeta(\vk_1-\vp_1) }
  {}
  { \zeta(\vp_2) }
\bcontraction[2ex]
  { \Big\langle }
  { \zeta(\vp_1) }
  {\zeta(\vk_1-\vp_1) \zeta(\vp_2) }
  { \zeta(\vk_2-\vp_2) }
\Big\langle \zeta(\vp_1) \zeta(\vk_1-\vp_1) \zeta(\vp_2) \zeta(\vk_2-\vp_2) \Big\rangle \\
(ii) \qquad & \bcontraction
  { \Big\langle }
  { \zeta(\vp_1) }
  { \zeta(\vk_1-\vp_1) }
  { \zeta(\vp_2) }
\bcontraction[2ex]
  { \Big\langle \zeta(\vp_1) }
  {\zeta(\vk_1-\vp_1) }
  { \zeta(\vp_2) }
  { \zeta(\vk_2-\vp_2) }
\Big\langle \zeta(\vp_1) \zeta(\vk_1-\vp_1) \zeta(\vp_2) \zeta(\vk_2-\vp_2) \Big\rangle & \text{(obtained from $(i)$ by $\vp_2\to (\vk_2-\vp_2)$ )}\\
\end{aligned}
\label{eq: contractions two-point}
\end{equation}
The two contractions $(i),\, (ii)$ correspond to the configuration of momenta shown in Fig.~\ref{fig: contractions two-point}.
The sum of the contractions $(i)$ and $(ii)$ gives
\begin{multline}
\Big\langle \zeta(\vp_1) \zeta(\vk_1-\vp_1) \zeta(\vp_2) \zeta(\vk_2-\vp_2) \Big\rangle = \\
  = (2\pi)^6 \vdelta(\vk_1+\vk_2) \Big[ \vdelta(\vk_2+\vp_1-\vp_2) +\vdelta(\vp_1+\vp_2) \Big]
  \frac{2\pi^2}{p_1^3} \frac{2\pi^2}{|\vk_1-\vp_1|^3}
  \Pz(p_1) \Pz(|\vk_1-\vp_1|).
\label{eq: contraction two-point final}
\end{multline}
The two contributions give the same result, given that they correspond to each other up to a shift $\vp_2\to (\vk_2-\vp_2)$, which is a symmetry of Eq.~\eqref{eq: two-pt h def}.
\begin{framed}{ \footnotesize \noindent 
To check the symmetry of the whole integral under the exchange of $\vp, \vk-\vp$, it is important to observe that, for a generic function $f$,
\begin{equation}
\int \mathrm d^3p\, e^{s,ij}(\vk)p_i p_j f(\vk-\vp) f(\vp) =
\int \mathrm d^3 \widetilde p \,e^{s,ij}(\vk)(k_i-\widetilde p_i)(k_j-\widetilde p_j) f(\widetilde{\vp}) f(\vk-\widetilde{\vp})=
\int \mathrm d^3 \widetilde p\, e^{s,ij}(\vk)\widetilde p_i\ \widetilde p_j f(\widetilde{\vp}) f(\vk-\widetilde{\vp}) ,
\label{eq: symmetry p to k-p}
\end{equation}
since $e^{s,ij}(\vk)$ is transverse to $\vk$.
}\end{framed}
\noindent
We can evaluate then Eq.~\eqref{eq: two-pt h def} for any of the two configurations, and multiply the final result by 2, to get Eq.~(\ref{eq: two-pt h contracted}) after integration over $\vp_2$ 
with a Dirac delta so that $\vp_2=\vp_1-\vk_1$, and $\vk_2=-\vk_1$.

\subsection{Six-point function of the curvature perturbation}
\noindent
To calculate the six-point function of $\zeta$ that appears in
\eqref{eq: 3-pt h def} we have eight possible contractions for $\vk_i\neq 0$, listed in Eq.~\eqref{eq: contractions 3-point}. 
This total number of eight can be understood as the product of four choices for the contraction of $\zeta(\vp_1)$ times the number of contractions for the remaining four $\zeta$'s, that is two. 
All these contractions yield the same contribution to the bispectrum, thanks to the invariance of Eq.~\eqref{eq: 3-pt h def} under the exchange of the subscripts $1$ and $2$ and under $\vp_i\to \vk_i-\vp_i$, as shown in Eq.~\eqref{eq: symmetry p to k-p} and \eqref{eq: Ic, Is}. 
\begin{align}
(i) \quad  & 
\bcontraction[1ex]
  { \Big\langle \zeta(\vp_1) }
  { \zeta(\vk_1-\vp_1) }
  { }
  { \zeta(\vp_2) }
\bcontraction[1ex]
  { \Big\langle \zeta(\vp_1) \zeta(\vk_1-\vp_1) \zeta(\vp_2) }
  { \zeta(\vk_2-\vp_2) }
  {  }
  { \zeta(\vp_3) }
\bcontraction[2ex]
  { \Big\langle }
  { \zeta(\vp_1) }
  { \zeta(\vk_1-\vp_1) \zeta(\vp_2) \zeta(\vk_2-\vp_2) \zeta(\vp_3) }
  { \zeta(\vk_3-\vp_3) }
\Big\langle \zeta(\vp_1) \zeta(\vk_1-\vp_1) \zeta(\vp_2) \zeta(\vk_2-\vp_2) \zeta(\vp_3) \zeta(\vk_3-\vp_3) \Big\rangle 
\label{eq: contractions 3-point}
\\
(ii) \quad  & 
\bcontraction[1ex]
  { \Big\langle }
  { \zeta(\vp_1) }
  { \zeta(\vk_1-\vp_1) }
  { \zeta(\vp_2) }
\bcontraction[1ex]
  { \Big\langle \zeta(\vp_1) \zeta(\vk_1-\vp_1) \zeta(\vp_2) }
  { \zeta(\vk_2-\vp_2) }
  {  }
  { \zeta(\vp_3) }
\bcontraction[2ex]
  { \Big\langle \zeta(\vp_1) }
  { \zeta(\vk_1-\vp_1) }
  { \zeta(\vp_2) \zeta(\vk_2-\vp_2) \zeta(\vp_3) }
  { \zeta(\vk_3-\vp_3) }
\Big\langle \zeta(\vp_1) \zeta(\vk_1-\vp_1) \zeta(\vp_2) \zeta(\vk_2-\vp_2) \zeta(\vp_3) \zeta(\vk_3-\vp_3) \Big\rangle 
  & \text{(obtained from $(i)$ by $\vp_1\to (\vk_1-\vp_1)$ )} \notag \\
(iii) \quad  & 
\bcontraction[2ex]
  { \Big\langle }
  { \zeta(\vp_1) }
  { \zeta(\vk_1-\vp_1) \zeta(\vp_2) \zeta(\vk_2-\vp_2)  }
  { \zeta(\vp_3) }
\bcontraction[1ex]
  { \Big\langle \zeta(\vp_1) }
  { \zeta(\vk_1-\vp_1) }
  {  }
  { \zeta(\vp_2) }
\bcontraction[1ex]
  { \Big\langle \zeta(\vp_1) \zeta(\vk_1-\vp_1) \zeta(\vp_2) }
  { \zeta(\vk_2-\vp_2) }
  { \zeta(\vp_3) }
  { \zeta(\vk_3-\vp_3) }
\Big\langle \zeta(\vp_1) \zeta(\vk_1-\vp_1) \zeta(\vp_2) \zeta(\vk_2-\vp_2) \zeta(\vp_3) \zeta(\vk_3-\vp_3) \Big\rangle 
  & \text{(obtained from $(i)$ by $\vp_3\to (\vk_3-\vp_3)$ )} \notag \\
(iv) \quad  & 
\bcontraction[3ex]
  { \Big\langle }
  { \zeta(\vp_1) }
  { \zeta(\vk_1-\vp_1) \zeta(\vp_2) \zeta(\vk_2-\vp_2) \zeta(\vp_3) }
  { \zeta(\vk_3-\vp_3) }
\bcontraction[1ex]
  { \Big\langle \zeta(\vp_1) }
  { \zeta(\vk_1-\vp_1) }
  { \zeta(\vp_2) }
  { \zeta(\vk_2-\vp_2) }
\bcontraction[2ex]
  { \Big\langle \zeta(\vp_1) \zeta(\vk_1-\vp_1) }
  { \zeta(\vp_2) }
  { \zeta(\vk_2-\vp_2) }
  { \zeta(\vp_3) }
\Big\langle \zeta(\vp_1) \zeta(\vk_1-\vp_1) \zeta(\vp_2) \zeta(\vk_2-\vp_2) \zeta(\vp_3) \zeta(\vk_3-\vp_3) \Big\rangle 
  & \text{(obtained from $(i)$ by $\vp_2\to (\vk_2-\vp_2)$ )} \notag \\  
(v) \quad  & 
\bcontraction[1ex]
  { \Big\langle }
  { \zeta(\vp_1) }
  { \zeta(\vk_1-\vp_1) \zeta(\vp_2) }
  { \zeta(\vk_2-\vp_2) }
\bcontraction[2ex]
  { \Big\langle \zeta(\vp_1) }
  { \zeta(\vk_1-\vp_1) }
  { \zeta(\vp_2) \zeta(\vk_2-\vp_2) }
  { \zeta(\vp_3) }
\bcontraction[3ex]
  { \Big\langle \zeta(\vp_1) \zeta(\vk_1-\vp_1) }
  { \zeta(\vp_2) }
  { \zeta(\vk_2-\vp_2) \zeta(\vp_3) }
  { \zeta(\vk_3-\vp_3) }
\Big\langle \zeta(\vp_1) \zeta(\vk_1-\vp_1) \zeta(\vp_2) \zeta(\vk_2-\vp_2) \zeta(\vp_3) \zeta(\vk_3-\vp_3) \Big\rangle   
  & \text{(obtained from $(i)$ by $1\leftrightarrow 2$ )} \notag \\  
(vi) \quad  & 
\bcontraction[1ex]
  { \Big\langle }
  { \zeta(\vp_1) }
  { \zeta(\vk_1-\vp_1) }
  { \zeta(\vp_2) }
\bcontraction[2ex]
  { \Big\langle \zeta(\vp_1) }
  { \zeta(\vk_1-\vp_1) }
  { \zeta(\vp_2) \zeta(\vk_2-\vp_2) }
  { \zeta(\vp_3) }
\bcontraction[1ex]
  { \Big\langle \zeta(\vp_1) \zeta(\vk_1-\vp_1) \zeta(\vp_2) }
  { \zeta(\vk_2-\vp_2) }
  { \zeta(\vp_3) }
  { \zeta(\vk_3-\vp_3) }
\Big\langle \zeta(\vp_1) \zeta(\vk_1-\vp_1) \zeta(\vp_2) \zeta(\vk_2-\vp_2) \zeta(\vp_3) \zeta(\vk_3-\vp_3) \Big\rangle   
  & \text{(obtained from $(i)$ by $\vp_1\to (\vk_1-\vp_1)$ and $1\leftrightarrow 2$ )} \notag \\
(vii) \quad  & 
\bcontraction[1ex]
  { \Big\langle }
  { \zeta(\vp_1) }
  { \zeta(\vk_1-\vp_1) \zeta(\vp_2) }
  { \zeta(\vk_2-\vp_2) }
\bcontraction[2ex]
  { \Big\langle \zeta(\vp_1) \zeta(\vk_1-\vp_1) }
  { \zeta(\vp_2) }
  { \zeta(\vk_2-\vp_2) }
  { \zeta(\vp_3) }
\bcontraction[3ex]
  { \Big\langle \zeta(\vp_1) }
  { \zeta(\vk_1-\vp_1) }
  { \zeta(\vp_2) \zeta(\vk_2-\vp_2) \zeta(\vp_3) }
  { \zeta(\vk_3-\vp_3) }
\Big\langle \zeta(\vp_1) \zeta(\vk_1-\vp_1) \zeta(\vp_2) \zeta(\vk_2-\vp_2) \zeta(\vp_3) \zeta(\vk_3-\vp_3) \Big\rangle   
  & \text{(obtained from $(i)$ by $\vp_3\to (\vk_3-\vp_3)$ and $1\leftrightarrow 2$ )} \notag \\
(viii) \quad  & 
\bcontraction[2ex]
  { \Big\langle }
  { \zeta(\vp_1) }
  { \zeta(\vk_1-\vp_1) \zeta(\vp_2) \zeta(\vk_2-\vp_2) }
  { \zeta(\vp_3) }
\bcontraction[1ex]
  { \Big\langle \zeta(\vp_1) }
  { \zeta(\vk_1-\vp_1) }
  { \zeta(\vp_2) }
  { \zeta(\vk_2-\vp_2) }
\bcontraction[3ex]
  { \Big\langle \zeta(\vp_1) \zeta(\vk_1-\vp_1) }
  { \zeta(\vp_2) }
  { \zeta(\vk_2-\vp_2) \zeta(\vp_3) }
  { \zeta(\vk_3-\vp_3) }
\Big\langle \zeta(\vp_1) \zeta(\vk_1-\vp_1) \zeta(\vp_2) \zeta(\vk_2-\vp_2) \zeta(\vp_3) \zeta(\vk_3-\vp_3) \Big\rangle   
  & \text{(obtained from $(i)$ by $\vp_2\to (\vk_2-\vp_2)$ and $1\leftrightarrow 2$ )} \notag 
\end{align}

\begin{figure}[h!] \centering
\raisebox{-\height}{$\hbox{ \convertMPtoPDF{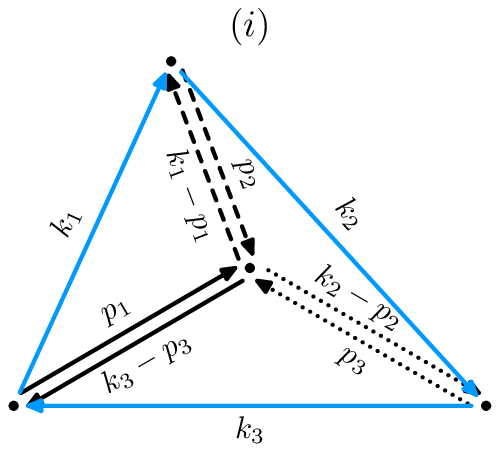}{\trA}{\trA} }$}
\raisebox{-\height}{$\hbox{ \convertMPtoPDF{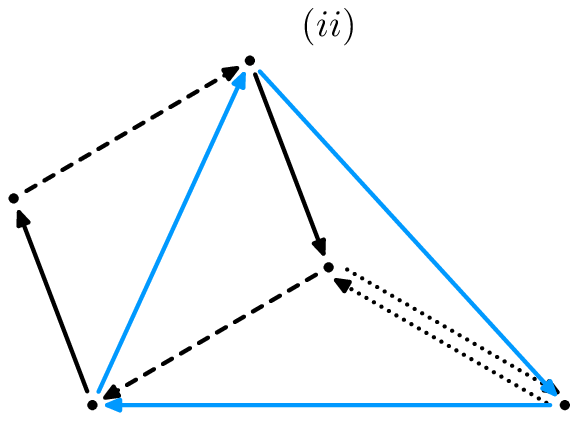}{\trA}{\trA} }$}
\raisebox{-\height}{$\hbox{ \convertMPtoPDF{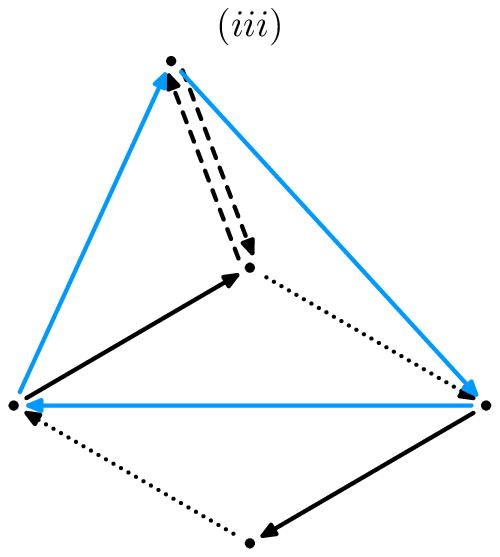}{\trA}{\trA} }$}
\raisebox{-\height}{$\hbox{ \convertMPtoPDF{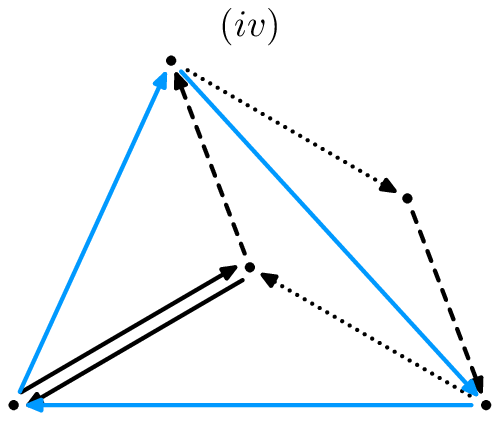}{\trA}{\trA} }$}
\raisebox{-\height}{$\hbox{ \convertMPtoPDF{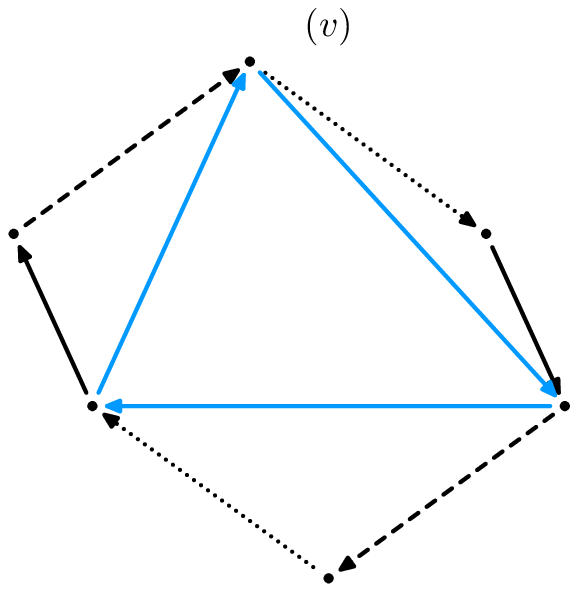}{\trB}{\trB} }$}
\raisebox{-\height}{$\hbox{ \convertMPtoPDF{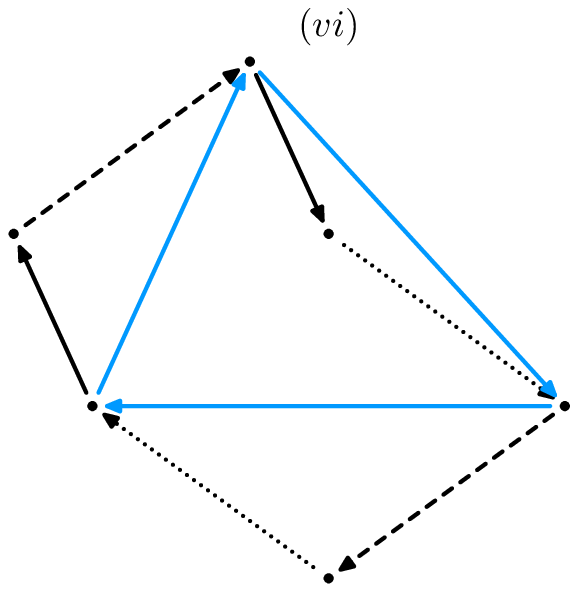}{\trB}{\trB} }$}
\raisebox{-\height}{$\hbox{ \convertMPtoPDF{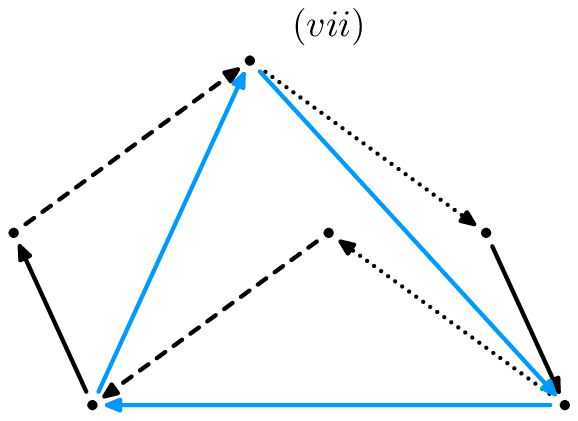}{\trB}{\trB} }$}
\raisebox{-\height}{$\hbox{ \convertMPtoPDF{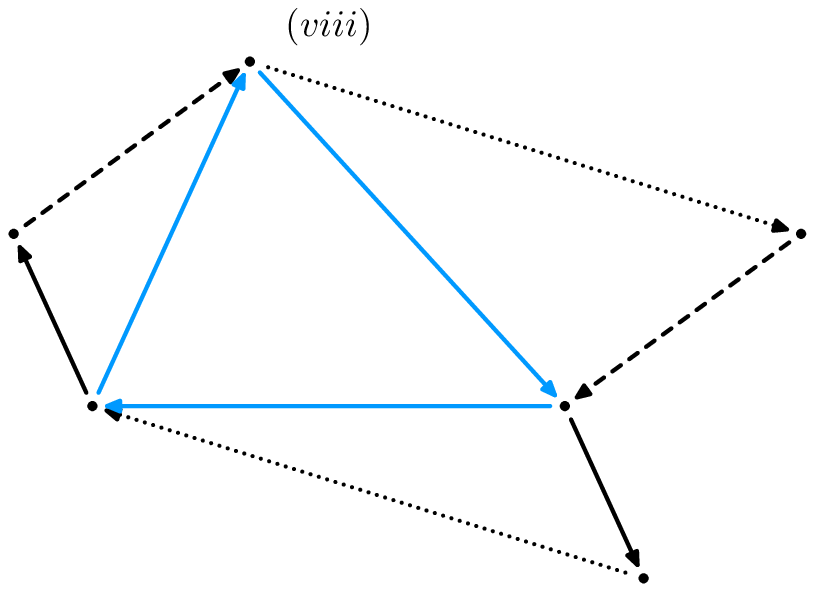}{\trB}{\trB} }$}
\caption{Geometrical configurations for the eight non-vanishing contractions of the three-point function listed in Eq.~\eqref{eq: contractions 3-point}. }
\label{fig: contractions three-point}
\end{figure}

In Fig.~\ref{fig: contractions three-point} we show the resulting geometrical configurations for the six momenta $\vp_i$, $(\vk_i-\vp_i)$, projected on the plane of the triangle formed by the $\vk_i$. Notice indeed that all the contractions result in a common factor $\vdelta(\vk_1+\vk_2+\vk_3)$. 
The labels of the vectors are printed only for the contraction $(i)$ to facilitate the reading. 
\newline
We can evaluate the three-point function for any of these configurations and multiply by eight the result. We choose the contraction $(i)$, which is equal to 
\begin{multline}
\bcontraction[1ex]
  { \Big\langle \zeta(\vp_1) }
  { \zeta(\vk_1-\vp_1) }
  { }
  { \zeta(\vp_2) }
\bcontraction[1ex]
  { \Big\langle \zeta(\vp_1) \zeta(\vk_1-\vp_1) \zeta(\vp_2) }
  { \zeta(\vk_2-\vp_2) }
  {  }
  { \zeta(\vp_3) }
\bcontraction[2ex]
  { \Big\langle }
  { \zeta(\vp_1) }
  { \zeta(\vk_1-\vp_1) \zeta(\vp_2) \zeta(\vk_2-\vp_2) \zeta(\vp_3) }
  { \zeta(\vk_3-\vp_3) }
\Big\langle \zeta(\vp_1) \zeta(\vk_1-\vp_1) \zeta(\vp_2) \zeta(\vk_2-\vp_2) \zeta(\vp_3) \zeta(\vk_3-\vp_3) \Big\rangle = \\
(2\pi)^9 \vdelta(\vk_1+\vk_2+\vk_3) \left(2\pi^2\right)^3 
 \frac{\Pz(p_1)}{p_1^3} \frac{\Pz(p_2)}{p_2^3} \frac{\Pz(p_3)}{p_3^3}
 \vdelta(\vp_1+\vk_3-\vp_3) \vdelta(\vk_1-\vp_1+\vp_2).
\label{eq: contraction 3pt (i)}
\end{multline}
We then proceed to the integration of the three-point function over the conjugate momenta.
The Dirac deltas in Eq.~\eqref{eq: contraction 3pt (i)} fix the geometrical configuration of the six momenta $\vk_i$, $\vp_i$ as shown in Fig.~\ref{fig: tetrahedron}.
%It is convenient, for the purpose of the numerical integration, to exploit the symmetry of the polarisation tensors under $\vp_i \to \alpha \vk_i-\vp_i$, as shown in Eq.~\eqref{eq: symmetry p to k-p}.
%The best choice is to define $\tvp_i\equiv \vk_i/two-\vp_i$, which enables us to use the Thales's theorem to write the simple relations
%\begin{equation}
%\tvp_1-\tvp_2 = \frac{\vk_3}{2} , \quad \tvp_three-\tvp_1 = \frac{\vk_1}{2} .
%\label{eq: tilde p Thales}
%\end{equation}
%
We can integrate over $\dd^3p_2$ and $\dd^3p_3$ in Eq.~\eqref{eq: 3-pt h def} with the last two Dirac deltas in \eqref{eq: contraction 3pt (i)}. The result is (\ref{eq: 3-pt h contracted}) and the remaining integral in $\dd^3 p_1$ has to be evaluated numerically.

\setcounter{equation}{0}
%%%%%%%%%%%%%%%%%%%%%%%%%%%%%%%
\section{Appendix C: The final curvature perturbation in the radiation phase}
%%%%%%%%%%%%%%%%%%%%%%%%%%%%%%%
\renewcommand{\theequation}{C.\arabic{equation}}
\noindent
In this Appendix we follow the evolution of the perturbations during the  reheating phase, which we consider for simplicity to be instantaneous (happening for instance in hybrid models in which a heavy waterfall field releases its vacuum  energy providing a fast transition from inflation to radiation), and the subsequent radiation phase. 
To model the sudden transition from inflation to radiation one can imagine that the equations of motion of the Higgs and its perturbations possess a time-dependent term taking care of the fast appearance of the plasma correction to the potential under  the form of the $m_T^2 h^2$ term.
By continuity,  $h_{\rm c}(t_{\rm e})=h_{\rm c}(t_{\rm RH})$,  $\dot{h}_{\rm c}(t_{\rm e})=\dot{h}_{\rm c}(t_{\rm RH})$, $\delta h(t_{\rm e})=\delta h(t_{\rm RH})$, and
$\delta \dot{h}(t_{\rm e})=\delta \dot{h}(t_{\rm RH})$, where $t_{\rm e}$ is the time at the end of inflation and $t_{\rm RH}$ is the time at the beginning of reheating. 
Assuming a fast reheating essentially amounts to saying that $t_{\rm e}\simeq t_{\rm RH}$. 
Across this time boundary, energy is also conserved. At the end of inflation the energy density is
\be
\rho_{\rm e} = \rho_{\rm inf} + \rho_{h,{\rm e}}\ ,
\ee
with $\rho_{\rm inf}=3H^2m_P^2$ and $\rho_{h,{\rm e}}=\dot h_{{\rm c},{\rm e}}^2/2 +V_0(h_{{\rm c},{\rm e}})$, with $h_{{\rm c},{\rm e}}\equiv h_{\rm c}(t_{\rm e})$. During the instantaneous reheating, $ \rho_{\rm inf} $ is used up in reheating the plasma (populated through the inflaton decays). The total energy density at $t_{\rm RH}$ is
\be
\rho_{\rm RH} = \rho_{\rm pl} + \rho_{h,{\rm RH}}\ .
\ee
In the plasma rest-frame 
\be
\rho_{\rm pl}=\omega - P,
\ee where
$P$ is the plasma pressure (equal to minus the free-energy density)
and 
\be
\omega=T\frac{\partial P}{\partial T}
\ee
 is the enthalpy density. We also have $\rho_{h,{\rm RH}}=\dot h_{{\rm c},{\rm RH}}^2/2 +V_0(h_{{\rm c},{\rm RH}})$, with $h_{{\rm c},{\rm RH}}\equiv h_{\rm c}(t_{\rm RH})$. It is more convenient to arrange the splitting between plasma and Higgs background energies in a different way, by first separating a pure radiation part in $\rho_{\rm pl}$ by writing 
\be
P=P_\gamma-V_T(h_{\rm c},T)
\ee
 and 
 \be
 \omega=\omega_\gamma-T\frac{\partial V_T}{\partial T},
 \ee 
 where $P_\gamma=\pi^2g_* T^4/90$, $\omega_\gamma=2\pi^2g_*T^4/45$ and $V_T(h_{\rm c},T)$ is the field-dependent thermal contribution of the plasma to the Higgs potential. Then, we assign this potential term to the Higgs energy
density and write
\be
\rho_{\rm pl} = \frac{\pi^2}{30}g_* T^4 -T\frac{\partial V_T}{\partial T}\ ,\quad
\rho_{\rm h,{\rm RH}} = \frac12 \dot h_{{\rm c},{\rm RH}}^2 + V_0(h_{{\rm c},{\rm RH}}) + V_T(h_{{\rm c},{\rm RH}},T).
\ee
The reheating temperature can be obtained from $\rho_{\rm e}=\rho_{\rm RH}$,
which gives $T_{\rm RH}\simeq [90/(\pi^2g_*)]^{1/4}\sqrt{H m_P}$. The small fluctuations in the Higgs background cause small fluctuations in $T_{\rm RH}$. Using $V_T\simeq m_T^2 h_{\rm c}^2/2=\kappa  T^2h_{\rm c}^2/2$ we get $\delta T\simeq 15\kappa h_{\rm c}\delta h_{\rm c}/(2\pi^2g_* T)\ll \delta h$. By matching the fluctuations in the energy density 
across $t_{\rm e}\simeq t_{\rm RH}$, that is, $\delta \rho_{\rm e}=\delta\rho_{\rm RH}$, we obtain 
\be
\delta \rho_{h,{\rm e}}=\delta \rho_{\rm pl} + \delta\rho_{h,{\rm RH}} \ ,
\ee
or, more explicitly,
\begin{eqnarray}
\delta \left[\frac12 \dot h_{{\rm c},{\rm e}}^2+ V_0(h_{{\rm c},{\rm e}})\right]_h &=&\left[\frac{2\pi^2}{15}g_*T^3\delta T-2\kappa T^2 h_{{\rm c},{\rm RH}}\delta h_{\rm RH}\right]_{\rm pl} +
\delta \left[\frac12 \dot h_{{\rm c},{\rm RH}}^2+ V_0(h_{{\rm c},{\rm RH}})+\frac12\kappa T^2 h^2_{{\rm c},{\rm RH}}\right]_h\nonumber\\
&=&\delta \left[\frac12 \dot h_{{\rm c},{\rm RH}}^2+ V_0(h_{{\rm c},{\rm RH}})\right]_h,
\end{eqnarray}
where in the last equality we have used the result for $\delta T$ above, which leads to a cancellation of the $\kappa T^2 h_{\rm c}\delta h$ terms.

Leaving aside Hubble friction, the energy density of plasma and Higgs background field are not conserved separately. We can still split the energy conservation equation $\dot \rho_{\rm tot}=0$ in
a plasma and a Higgs one, taking into account Higgs decays into the plasma and write
\be
\dot \rho_h = (\partial \rho_h/\partial h_{\rm c}) \dot h_{\rm c} =(\Box h_c +V')\dot h_{\rm c}  = - \gamma_h \dot h_{\rm c} ^2 \ ,
\label{hEoMdecay}
\ee
with $V=V_0+V_T$, and  $\gamma_h\simeq 10^{-3} T$ the Higgs decay width, while $\dot \rho_{\rm pl}=+\gamma_h\dot h_{\rm c} ^2$. 
The right-hand side in Eq. (\ref{hEoMdecay}) introduces a friction term in the equation of motion for the Higgs field that is initially subleading in comparison with the Hubble friction term that it should also include, but is important for the late time behaviour of the Higgs condensate.

Having understood how to properly deal with the perturbations of the Higgs coupled to the plasma at a given temperature $T$, we are now ready to deal with their behaviour in the radiation phase.
Introducing Hubble friction and neglecting the decay term we write the equations of motion of the classical value of the Higgs and its linear  perturbation still in the flat gauge as
\begin{eqnarray}
\ddot h_{\rm c}+\frac{3}{2 t}\dot h_{\rm c}+m_T^2 h_{\rm c}&=&0,\nonumber\\
\delta\ddot{h}_{1}+\frac{3}{2 t}\delta\dot{h}_{1}+m_T^2\delta  h_{1}&=&0.
%\delta\ddot{h}_{2}+\frac{3}{2 t}\delta\dot{h}_{2}+m_T^2\delta  h_{2}&=&0.
\end{eqnarray}
Notice that in  the   radiation phase the temperature scales as 
\be
T= T_{\rm RH}\left(\frac{a_{\rm RH}}{a}\right)
\ee 
and the scale factor follows the rule $a=a_{\rm RH}(t/t_{\rm RH})^{1/2}$, so that $H=1/(2t)$.
One finds
\be
h_{\rm c}(t)=  h(t_{\rm RH})\sqrt{\frac{t_{\rm RH}}{t}}{\rm cos}\left[2 m_{T_{\rm RH}}(\sqrt{t t_{\rm RH}}-t_{\rm RH})\right]+
\frac{2\dot h(t_{\rm RH}) t_{\rm RH}+h(t_{\rm RH})}{2m_{T_{\rm RH}} \sqrt{t t_{\rm RH}}}{\rm sin}\left[2 m_{T_{\rm RH}}(\sqrt{t t_{\rm RH}}-t_{\rm RH})\right],
\ee
and $\delta  h_{1}$  tracks $h_{\rm c}$.
Up to fast oscillations  the classical Higgs field scales like   
\be
h_{\rm c}(a)\sim \frac{1}{a}\sim T
\ee
and one can show easily that  averaging over the fast oscillations  one gets $\langle \dot h_{\rm c}^2\rangle=\langle m_T^2  h_{\rm c}^2\rangle$, so that
\be
\dot{\rho}_h=\dot h_{\rm c}\ddot h_{\rm c}+m_T^2  h_{\rm c}\dot h_{\rm c}+\frac{1}{2}\dot{m}^2_T h_{\rm c}^2=-3H \dot h_{\rm c}^2-Hm_T^2  h_{\rm c}^2,
\ee
and upon averaging one gets
\be
\dot{\rho}_h=-4H\rho_h.
\ee
The Higgs before decaying behaves therefore like a relativistic fluid.  One obtains also (still in the flat gauge)
\be
\rho_h=\rho_h+\delta\rho_{h,1}=m_T^2h_{\rm c}^2+2 m_T^2h_{\rm c}\delta h_1,
\ee
so that 
\be
\frac{\delta\rho_{h,1}}{\rho_h}=2\frac{\delta h_1(t)}{h_{\rm c}(t)}=2\frac{\delta h_1(t_{\rm RH})}{h_{\rm c}(t_{\rm RH})}=2\frac{\delta h_1(t_{\rm e})}{h_{\rm c}(t_{\rm e})},
\ee
where in the next-to-last passage we have used the fact that $\delta  h_{1}$ is tracking $h_{\rm c}$ and in the last passage the continuity in the Higgs sector.
We have therefore that upon Higgs decay and  using $\delta\dot{\rho}_{h,1}=-4H\delta\rho_{h,1}$,
\begin{eqnarray}
-\zeta_{1}(t_{\rm dec})&=&H\frac{\delta\rho_{1}(t_{\rm dec})}{\dot\rho(t_{\rm dec})}= Hr_h(t_{\rm dec})\frac{\delta\rho_{h,1}(t_{\rm dec})}{\dot\rho_{h}(t_{\rm dec})}=
-\frac{r_h(t_{\rm dec})}{4}\frac{\delta\rho_{h,1}(t_{\rm dec})}{\rho_h(t_{\rm dec})}\nonumber\\
&=&-\frac{r_h(t_{\rm dec})}{2}\frac{\delta h_1(t_{\rm RH})}{h_{\rm c}(t_{\rm RH})}=-\frac{r_h(t_{\rm dec})}{2}\frac{\delta h_1(t_{\rm e})}{h_{\rm c}(t_{\rm e})}=-\frac{r_h(t_{\rm dec})}{2}\frac{\dot{h}_{\rm c}(t_{\rm e})}{H h_{\rm c}(t_{\rm e})}\zeta_{h,1}(t_{\rm e}),
\end{eqnarray}
where we have made use of the fact that during inflation
\be
-\zeta_{h,1} = H\frac{\delta\rho_{h,1}}{\dot\rho_h}=H\frac{\delta h_{1}}{\dot{h}_{\rm c}}.
\ee
In particular, notice that $\zeta_h$ during inflation does not coincide with the value during reheating, $\zeta_h\simeq (\delta h_1/2{h}_{\rm c})$, signalling that $\zeta_h$ is not conserved during the transition. This is not surprising, as the Higgs interacts with the hot plasma to suddenly acquire a plasma mass and therefore is not an isolated fluid.
The  final power spectrum reads
\be
\label{c21}
{\cal P}_\zeta(t_{\rm dec})=\frac{k^3}{2\pi^2}|\zeta_k(t_{\rm dec})|^2=\frac{r^2_h(t_{\rm dec})}{4}\left(\frac{H}{2\pi}\right)^2\left(\frac{\dot{h}_{\rm c}(t_{\rm e})}{h_{\rm c}(t_{\rm e})\dot{h}_{\rm c}(t_k)} \right)^2.
\ee

\setcounter{equation}{0}
%%%%%%%%%%%%%%%%%%%%%%%%%%%%%%%
\section{Appendix D: Analytical results for the functions ${\bma \Ic}, {\bma \Is}$}
\label{sec: app Ic Is}
%%%%%%%%%%%%%%%%%%%%%%%%%%%%%%%
\renewcommand{\theequation}{D.\arabic{equation}}

To write down the analytical formul\ae\ for the integrals $\Ic, \Is$ defined in Eq.~\eqref{eq: Ic, Is} we use the indefinite integrals 
$i_c(d,s,\tau)$ and $i_s(d,s,\tau)$, so that $\Ic(d,s)=\left.i_c(d,s,\tau)\right|_1^\infty$ and a similar formula for $\Is$.
We get
\bea
i_c(d,s,\tau) &=& \frac{288}{(s^2-d^2)^3}
 \Big\{\left[A(s)\cos\tau+C(d,s)\sin\tau\right]\cos (d\tau)
- \left[A(d)\cos\tau+C(s,d)\sin\tau\right]\cos (s\tau)\nonumber\\
&+& \left[(2/\tau^2)\cos\tau+B(s)\sin\tau\right]d\sin (d\tau)
- \left[(2/\tau^2)\cos\tau+B(d)\sin\tau\right]s \sin (s\tau)\nonumber\\
& +&\frac18 (s^2+d^2-2)^2\left({\rm Si}[(1-s)\tau]+{\rm Si}[(1+s)\tau]-{\rm Si}[(1+d)\tau]-{\rm Si}[(1-d)\tau]\right)\Big\}\ ,
\eea
and
\bea
i_s(d,s,\tau) &=& \frac{288}{(s^2-d^2)^3}
 \Big\{\left[A(s)\sin\tau-C(d,s)\cos\tau\right]\cos (d\tau)
+ \left[-A(d)\sin\tau+C(s,d)\cos\tau\right]\cos (s\tau)\nonumber\\
&+& \left[(2/\tau^2)\sin\tau-B(s)\cos\tau\right] d\,\sin (d\tau)
+\left[-(2/\tau^2)\sin\tau+B(d)\cos\tau\right]s\sin (s\tau)\nonumber\\
& +&\frac18 (s^2+d^2-2)^2\left({\rm Ci}[(1+d)\tau]+{\rm Ci}[(1-d)\tau]-{\rm Ci}[|s-1|\tau]-{\rm Ci}[(1+s)\tau]\right)\Big\}\ ,
\eea
where ${\rm Si}(x)$ is the sine integral function, ${\rm Ci}(x)$ is the cosine integral function
and
\be
A(x)\equiv\frac{1}{\tau^3}\left[2+(x^2-1)\tau^2\right]\ ,\quad
B(x)\equiv\frac{1}{\tau^3}\left[6+(x^2-1)\tau^2\right]\ ,\quad
C(x,y)\equiv\frac{1}{\tau^4}\left[6-(1+2x^2-y^2)\tau^2\right]\ .
\ee
Noting that
\be
i_c(d,s,\infty)=-36\pi\frac{(s^2+d^2-2)^2}{(s^2-d^2)^3}\theta(s-1)\ ,\quad\quad
i_s(d,s,\infty)=0\ ,
\ee
we get
\begin{multline}
\label{eq: Ic tau0=1}
\Ic(d,s) = \frac{288}{(s^2-d^2)^3}
 \Big\{\left[2c_1+(5+d^2)s_1\right]s\sin s-\left[2c_1+(5+s^2)s_1\right]d\sin d \\
 + \left[(1+d^2)c_1+(5+d^2-2s^2)s_1\right]\cos s-
\left[(1+s^2)c_1+(5+s^2-2d^2)s_1\right]\cos d \\
 +\frac18 (s^2+d^2-2)^2\left[{\rm Si}(1+d)+{\rm Si}(1-d)-{\rm Si}(1-s)-{\rm Si}(1+s)-\pi\theta(s-1)\right]\Big\}\ ,
\end{multline}
where $c_1\equiv \cos(1)\simeq 0.54$, $s_1\equiv \sin(1)\simeq 0.84$ and
\begin{multline}
\label{eq: Is tau0=1}
\Is(d,s) = \frac{288}{(s^2-d^2)^3}\Big\{\left[2s_1-(5+d^2)c_1\right]s\sin s+\left[-2s_1+(5+s^2)c_1\right]d\sin d \\
 + \left[(1+d^2)s_1-(5+d^2-2s^2)c_1\right]\cos s+
\left[-(1+s^2)s_1+(5+s^2-2d^2)c_1\right]\cos d \\
 - \frac18 (s^2+d^2-2)^2\left[{\rm Ci}(1+d)+{\rm Ci}(1-d)-{\rm Ci}(|1-s|)-{\rm Ci}(1+s)\right]\Big\}\ .
\end{multline}
If instead of the lower integration limit $\tau_0=1$ one takes $\tau_0=0$, using
\be
i_c(d,s,0)=0\ ,\quad\quad
i_s(d,s,0)=36\frac{(s^2+d^2-2)}{(s^2-d^2)^2}\left[\frac{(s^2+d^2-2)}{(s^2-d^2)}\log\frac{(1-d^2)}{|s^2-1|}+2\right]\ ,
\ee
one gets
\begin{equation}
\label{eq: Ic, Is tau0=0}
\begin{aligned}
\hat\Ic(d,s)&\equiv \left. i_c(d,s,\tau)\right|_0^\infty=-36\pi\frac{(s^2+d^2-2)^2}{(s^2-d^2)^3}\theta(s-1)\ ,\\
\hat\Is(d,s)&\equiv \left. i_s(d,s,\tau)\right|_0^\infty=-36\frac{(s^2+d^2-2)}{(s^2-d^2)^2}\left[\frac{(s^2+d^2-2)}{(s^2-d^2)}\log\frac{(1-d^2)}{|s^2-1|}+2\right]\ .
\end{aligned}
\end{equation}

All our results for the sourced gravitational waves and their power spectrum are in agreement with Ref.~\cite{Kohri:2018awv}. The only difference is in the lower limit of integration, which they have taken to be $\tau_0=0$ as in Eq.~\eqref{eq: Ic, Is tau0=0}. 
In our numerical results in this paper, we have instead chosen $\tau_0=1$ as in Eqs.~\eqref{eq: Ic tau0=1} and \eqref{eq: Is tau0=1}.
Strictly speaking, $\tau_0=0$ is the correct choice, but given that the source is damped on super-Hubble scales, the time integral receives the main contribution for $\tau\gtrsim 1$, and the final numerical difference is not particularly important.\\
We have also compared our results with \cite{Ananda:2006af}, finding some differences. 
In particular, their formul\ae\ for the power spectrum of GWs differ from ours by a total prefactor of $\frac{16}{81}\frac{32 k^2}{\pi^4}=0.06k^2$ that they introduced throughout their derivation, the sum over $k$ in their Eq.~(26) should start from $k=0$ and a few of the entries of the matrices $M^i_{mn}$ contain some typos.

\end{appendices}

%%%%%%%%%%%%%%%%%%%%%%%%%%%%%%%%%%%%%%%%%%

\end{document}